\documentclass[fleqn,usenatbib]{mnras}

\usepackage{newtxtext,newtxmath}

\usepackage[T1]{fontenc}

\DeclareRobustCommand{\VAN}[3]{#2}
\let\VANthebibliography\thebibliography
\def\thebibliography{\DeclareRobustCommand{\VAN}[3]{##3}\VANthebibliography}

\usepackage{graphicx}
\usepackage{amsmath}
\usepackage{caption}

\newcommand{\lya}{\mbox{Ly\,{\sc $\alpha$}}}
\newcommand{\hi}{\mbox{H\,{\sc i}}}
\newcommand{\hii}{\mbox{H\,{\sc ii}}}

\newcommand{\heii}{\mbox{He\,{\sc ii}}}
\newcommand{\heiiuv}{\mbox{He\,{\sc ii}{$\lambda 1640$}}}
\newcommand{\heiiopt}{\mbox{He\,{\sc ii}{$\lambda 4686$}}}

\newcommand{\oiii}{\mbox{[O\,{\sc iii]}}}

\newcommand{\oii}{\mbox{[O\,{\sc ii]}{$\lambda 3727$}}}
\newcommand{\oiinwl}{\mbox{[O\,{\sc ii]}}}

\newcommand{\oiiia}{\mbox{[O\,{\sc iii]}{$\lambda 5007$}}}
\newcommand{\oiiib}{\mbox{[O\,{\sc iii]}{$\lambda 4960$}}}
\newcommand{\oiiiaur}{\mbox{[O\,{\sc iii]}{$\lambda 4363$}}}

\newcommand{\halpha}{\mbox{H\,{\sc $\alpha$}}}
\newcommand{\hbeta}{\mbox{H\,{\sc $\beta$}}}
\newcommand{\hgamma}{\mbox{H\,{\sc $\gamma$}}}
\newcommand{\hdelta}{\mbox{H\,{\sc $\delta$}}}
\newcommand{\heps}{\mbox{H\,{\sc $\epsilon$}}}

\newcommand{\neiiia}{\mbox{[Ne\,{\sc iii]}{$\lambda 3869$}}}
\newcommand{\neiiib}{\mbox{[Ne\,{\sc iii]}{$\lambda 3967$}}}

\usepackage{orcidlink}

\title[An extremely metal-poor galaxy at $z=8.271$]{The \emph{JWST} EXCELS survey: an extremely metal-poor galaxy at $\mathbf{z=8.271}$ hosting an unusual population of massive stars}

\author[F. Cullen et al.]{F. Cullen\orcidlink{0000-0002-3736-476X}$^{1}$\thanks{E-mail: fergus.cullen@ed.ac.uk},
A. C. Carnall\orcidlink{0000-0002-1482-5818}$^{1}$,
D. Scholte\orcidlink{0000-0002-6867-1244}$^{1}$,
D.\,J. McLeod\orcidlink{0000-0003-4368-3326}$^{1}$,
R. J. McLure$^{1}$,
K. Z. Arellano-C\'ordova\orcidlink{0000-0002-2644-3518}$^{1}$,\and
T. M. Stanton\orcidlink{0000-0002-0827-9769}$^{1}$,
C. T. Donnan\orcidlink{0000-0002-7622-0208}$^{1}$,
J. S. Dunlop\orcidlink{0000-0002-1404-5950}$^{1}$,
A. E. Shapley,$^{2}$
L. Barrufet$^{1}$,
R. Begley$^{1}$,
C. Bondestam$^{1}$,\and
M. Cirasuolo$^{3}$,
H.-H. Leung$^{1}$, 
C. L. Pollock$^{4}$,
S. Stevenson$^{1}$
\footnotesize\\\\
$^{1}$Institute for Astronomy, University of Edinburgh, Royal Observatory, Edinburgh, EH9 3HJ, UK\\
$^{2}$Department of Physics \& Astronomy, University of California, 430 Portola Plaza, Los Angeles CA 90095, USA\\
$^{3}$European Southern Observatory, Karl-Schwarzschild-Strasse 2, D-85748 Garching bei Muenchen, Germany\\
$^{4}$The Cosmic Dawn Center, Niels Bohr Institute, University of Copenhagen, Jagtvej 128, DK-2200 Copenhagen N, Denmark
}

\date{Accepted 2025 May 12. Received 2025 April 29; in original form 2025 Jan 13\vspace{-0.2cm}}

\pubyear{2022}

\begin{document}
\label{firstpage}
\pagerange{\pageref{firstpage}--\pageref{lastpage}}
\maketitle

\begin{abstract}
We present an analysis of the rest-frame optical ($\lambda \simeq 3100-5600 \,${\AA}) spectrum of a $\mathrm{log}_{10}(M_*/\mathrm{M_\odot}) = 8.6$ star-forming galaxy at $z=8.271$ from \emph{JWST}/NIRSpec medium-resolution observations taken as part of the Early eXtragalactic Continuum and Emission Line Science (EXCELS) survey.
The galaxy (EXCELS-63107) is compact, with a size consistent with local star-forming cluster complexes ($r_e < 200 \, \rm{pc}$), and exhibits an extremely steep UV continuum slope measured from \emph{JWST}/NIRCam photometry ($\beta=-3.3\pm0.3$).
The \emph{JWST}/NIRSpec G395M spectrum of EXCELS-63107 is notable for its strong \oiiiaur \ auroral-line emission relative to the \oiiia \ forbidden line.
Via a detailed emission-line and photoionization-modelling analysis, we find that the observed properties of EXCELS-63107 are consistent with the presence of an ionizing source with an effective temperature of $T_{\rm eff} \gtrsim 80 \, 000\,\rm{K}$  heating ionized gas with a density of $n_e < 10^4 \, \rm{cm}^{-3}$ to a volume-averaged electron temperature of $T_e \simeq 34 \, 000\,\rm{K}$.
Crucially, we find that stellar population models assuming a standard IMF are not capable of producing the required heating.
We determine an oxygen abundance of ${12+\mathrm{log(O/H)}= 6.89^{+0.26}_{-0.21}}$ ($\simeq 1.6$ percent of solar) which is one of the lowest directly constrained oxygen abundances measured in any galaxy to date, and $\simeq 10 \times$ lower than is typical for $z\simeq8$ galaxies with the same stellar mass.
The extremely low metallicity of EXCELS-63107 places it in a regime in which theoretical models expect a transition to a top-heavy IMF, and we speculate that a $\simeq 10-30 \, \times$ excess of $M > 50 \, \rm{M}_{\odot}$ stars is one plausible explanation for its observed properties.
However, more exotic scenarios, such as Pop III star formation within a mildly enriched halo, are also consistent with the observations.

\end{abstract}

\begin{keywords}
galaxies: evolution - galaxies: formation - galaxies: high-redshift - galaxies: abundances - galaxies: starburst -  dark ages, reionization, first stars
\vspace{-0.5cm}
\end{keywords}



\section{Introduction}
\label{sec:intro}

The most metal-poor stellar populations at high redshift are important probes of the earliest phases of star formation and chemical enrichment, providing key constraints on the physical conditions and processes that governed the formation of the first galaxies.
Initial estimates of galaxy chemical abundances at $z \gtrsim 8$ with \emph{JWST} have indicated a perhaps surprising level of early enrichment, with typical metallicities (primarily traced via the gas-phase oxygen abundance) of approximately 10 per cent of the solar value ($Z \simeq 0.1 \, \rm{Z}_{\odot}$; e.g. \citealp{arellano-cordova-2024-ers-abundances-z78, curti_2023_ers_oh, langeroodi_2023_mzr_jwst, nakajima_2023_z410_mzr, trump_2023_ers_oh, curti_2024_jades_mzr, sanders_2024_jwst_highz_te, hsiao_2024_z10_oh_te, sarkar_2025_jwst_mzr}).
Even the highest-redshift galaxy currently spectroscopically confirmed ($z \simeq 14.2$) displays prominent highly-ionized oxygen species and is estimated to have ${Z \simeq 0.1-0.2 \, \rm{Z}_{\odot}}$ (\citealp{carniani_2024_z14, canrniani_2024_z14_metal, schouws_2024_z14_metal}).

However, it is clear from both the stellar archaeological record, and high-redshift damped \lya \ absorption systems, that stellar populations must have formed in environments with ${Z << 0.1\,\rm{Z}_{\odot}}$ at early times \citep[e.g.][]{cooke_2011_dla_low_metal, frebel_2015_xmp_stars, kobayashi_2020_gce_model}.
Additionally, in the local Universe, dwarf galaxies with metallicities as low as ${Z \simeq 0.01 \, \rm{Z}_{\odot}}$ have been discovered \citep[e.g.][]{kojima_2020_empgs, isobe_2022_local_xmpgs}.
Although promising candidates for extremely metal-poor, perhaps even metal-free, objects have been identified in some high-redshift studies \citep{maiolino_2024_pop3_gnz11, vanzella_pop3_like_cluster, willott_2025_xmpg_canucs}, these have so far been based on empirical diagnostics and as of yet, very few galaxies have been discovered with $Z \lesssim 0.01 \, \rm{Z}_{\odot}$ via the direct temperature-based approach (with rare exceptions, e.g. \citealp{laseter_2024_direct_oh_jades}).
After two years of \emph{JWST} operations, it remains true that the majority of the known extremely metal-poor galaxies are dwarf galaxies in the local Universe.
Uncovering and characterising the $Z \lesssim 0.01 \, \rm{Z}_{\odot}$ population that must exist at high redshifts therefore remains an open problem.

Metal-poor galaxies at high redshift are not only important probes of early star formation and chemical enrichment, but also provide crucial insights into the properties of the low-metallicity massive stars that presumably dominated stellar feedback and drove the process of reionization in the early Universe.
Individual massive metal-poor stars are both rare and difficult to observe in the local Universe \citep[e.g.][]{telford_2024_metalpoor_stars}, and therefore key properties such as stellar mass-loss rates and ionizing-continuum strengths have to be estimated using theoretical models \citep[e.g.][]{eldridge_2022_araa_review_massive_stars}.
One essential test of these theoretical stellar models is the ionized gas spectra of metal-poor galaxies in the rest-frame UV and optical, which encodes the properties of the ionizing stellar source.
In many instances, the high level of ionization observed in low-metallicity objects has proved challenging for standard stellar models to explain \citep[e.g.][]{olivier_2022_eelg_ionizing_spectrum, umeda_2022_eelg_ionizing_spectrum}.

In addition to the uncertainties related to the properties of individual metal-poor stars, the demographics of stellar populations in metal-poor environments also remain uncertain.
Several early \emph{JWST} studies have indeed reported indirect evidence for a top-heavy IMF in some high-redshift galaxies, with observational signatures including strong nebular continuum emission \cite[e.g.][]{cameron_2024_nebular_cont, katz_2024_balmerjumps} and anomalous chemical abundance patterns (particularly N-enrichment; \citealp{vink_2023_vms, bekki_2023_th_imf_no_ratios, curti_2024_z9_cno}).
Some theoretical models have also invoked a top-heavy IMF to explain the number density of UV-bright galaxies at $z>10$ \citep[e.g.][]{hutter_2024_th_imf_uvlf}.
Based on our theoretical understanding, an obvious avenue for investigating the IMF at high redshift is via detailed observations of the most metal-poor systems \citep[e.g.][]{mowla_2024_firefly}.

In this paper, we report the discovery and characterisation of a star-forming galaxy at $z=8.271$ (EXCELS-63107).
Via an analysis of a deep rest-frame optical \emph{JWST}/NIRSpec G395M spectrum, we determine a direct-method gas-phase metallicity of ${12+\mathrm{log(O/H)}= 6.89^{+0.26}_{-0.21}}$ ($Z \simeq 0.016 \, \rm{Z}_{\odot}$) which makes EXCELS-63107 one of the lowest-metallicity galaxies known at any redshift.
Interestingly, the spectrum of EXCELS-63107 also indicates the presence of extremely hot gas with an electron temperature of $T_e > 30 \, 000 \,\rm{K}$, but no obvious sign of AGN heating.
Via a detailed modelling analysis, we find that the properties of EXCELS-63107 cannot be explained by standard stellar models of massive stars and that alternative scenarios, such as a top-heavy IMF, need to be invoked.
More intriguingly, the observations are at least qualitatively consistent with more exotic explanations, such as pristine Pop III star formation occurring within a pre-enriched halo \citep[e.g.][]{venditti_2023_pop3_pop2_formation, correa_magnus_2024_pop3_formation_channel}.

The paper is organised as follows.
In Section~\ref{sec:obs_and_data_reduction} we describe our observations and data reduction.
In Section~\ref{sec:photometry_and_morphology} we present the photometric properties of EXCELS-63107 before discussing the features of its rest-frame optical spectrum in Section~\ref{sec:rest_optical_spectrum}.
Section~\ref{sec:metallicity_and_ionization} details the emission-line and photoionisation-modelling analyses that we use to determine the gas-phase metallicity and investigate the properties of the massive stellar population.
We discuss our results in Section~\ref{sec:discussion_of_excels63107} and summarise our findings in Section~\ref{sec:summary_and_conclusions}.
Throughout this paper, we refer to `standard' and `non-standard' stellar models as a convenient shorthand.
By `standard', we mean any commonly-used stellar population models such as \textsc{bpass} \citep{eldridge_2017_bpass}, \textsc{starburst99} \citep{leitherer_1999_starburst99, leithere_2014_starburst99}, \textsc{BC03} \citep{bc03_stellar_models} or Flexible Stellar Population Synthesis (\textsc{fsps}; \citealp{conroy_2009_fsps}), which assume a standard \citet{salpeter_imf} IMF high-mass slope and are for the most part successful in reproducing the observed properties of star-forming galaxies.
By `non-standard', we are specifically referring to changes in the IMF which produce an excess of massive stars, or more extreme stellar models that predict stronger ionizing continuum spectra. 
Throughout, we assume a standard cosmological model with $H_0=70$\,km s$^{-1}$ Mpc$^{-1}$, $\Omega_{\rm{m}}=0.3$ and $\Omega_{\Lambda}=0.7$ and a solar oxygen abundance of $\mathrm{12+log(O/H)} = 8.69$ \citep{asplund_2021_solar_abn}.

\begin{figure*}
        \centering
        \includegraphics[width=0.95\linewidth]{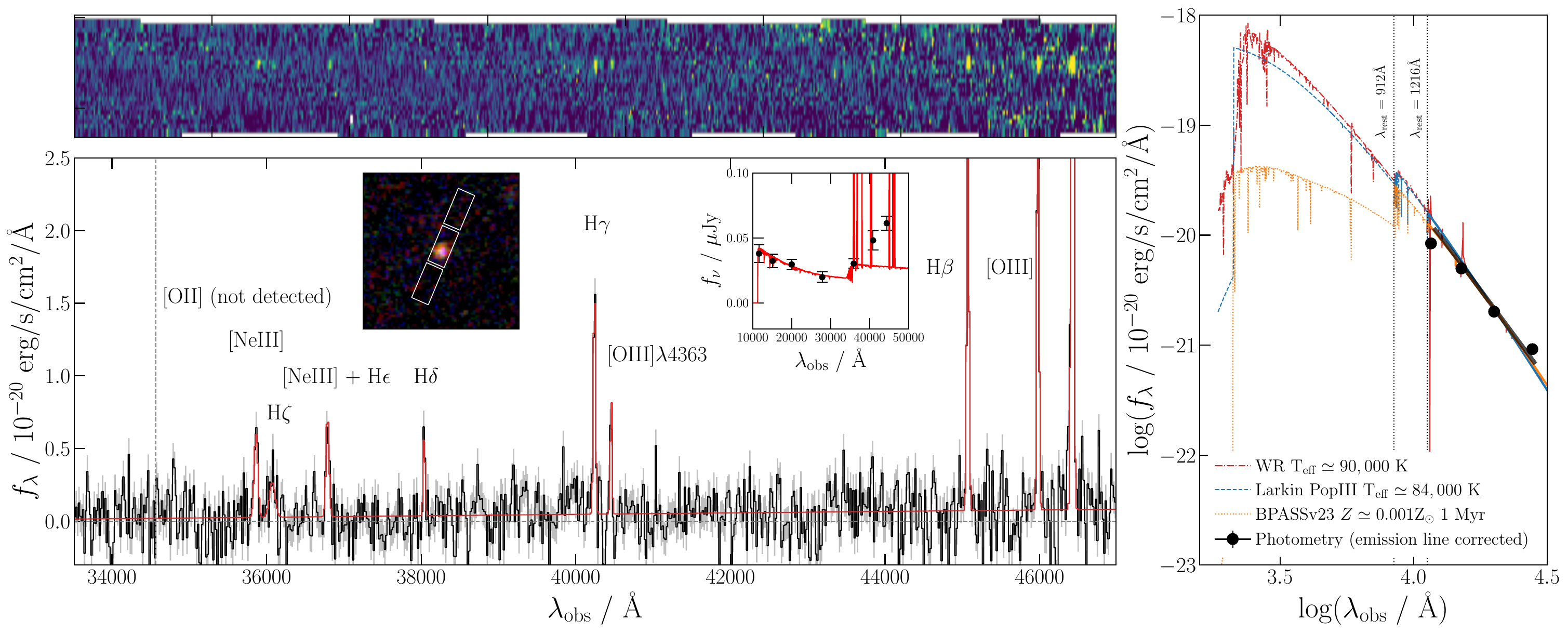}
        \caption{The \emph{JWST} NIRCam and NIRSpec data for EXCELS-63107.
        The upper and lower left-hand panels show the 2D and 1D spectrum for EXCELS-63107 respectively.
        The inset panels show a three-colour image [F200W, F365W, F444W] with the NIRSpec MSA shutter position overlaid (left-hand inset) and the full rest-frame UV to optical SED including the best-fitting \textsc{bagpipes} SED (right-hand inset).
        Notable spectral features include the prominent [O\,{\sc iii}]$\lambda$4363 and [Ne\,{\sc iii}]$\lambda$3869 emission lines and a distinct lack of the [O\,{\sc ii}]$\lambda$3726,29 doublet.
        These spectral features are best explained by a hot, low metallicity and highly ionized ISM.
        The right-hand panel shows rest-frame ultraviolet (UV) photometry from which we infer an extremely steep UV continuum slope of $\beta=-3.3\pm0.3$ (black line; where $f_{\lambda} \propto \lambda^{\beta}$).
        We also show the best-fitting SEDs for each of the three stellar population models types considered in our photoionization modelling analysis [\textsc{BPASSv2.3} (orange dotted line); Wolf-Rayet (blue dashed line); Pop III stars (red dot-dashed line); see Section \ref{subsec:cloudy_analysis}] highlighting the fact that the UV SED is consistent with pure stellar emission with an escape fraction close to unity (i.e. no nebular continuum emission).
        We show the extension of these SEDs beyond the Lyman break ($\lambda_{\rm{rest}} < 1216 \rm{\AA}$; dashed lines) to illustrate the significant difference in the ionizing flux.} 
        \label{fig:spec_and_sed}
    \end{figure*}

\section{Observations and Data Reduction}
\label{sec:obs_and_data_reduction}

The galaxy analysed in this work (EXCELS-63107) is taken from the \emph{JWST} Early eXtragalactic Continuum and Emission Line Science (EXCELS) survey (GO 3543; PIs: Carnall, Cullen; \citealp{carnall_2024_excels}).
The EXCELS survey is designed to provide deep, medium-resolution ($\mathrm{R}=1000$) spectroscopy across four NIRSpec Micro-Shitter Array (MSA) pointings within the \emph{JWST} Public Release IMaging for Extragalactic
Research (PRIMER) Ultra-Deep Survey (UDS) imaging footprint \citep{dunlop_2021_primer}.
The four pointings are observed with the G140M/F100LP, G235M/F170LP and G395M/F290LP gratings. 
Observations were conducted using a 3-shutter slitlet and 3-point dither pattern.
The exposure times in each grating were $\simeq 4$ hours in G140M, $\simeq 5.5$ hours in G235M and $\simeq 4$ hours in G395M.
The primary targets of the EXCELS survey were massive quiescent galaxies up to $z\simeq 5$ \citep{carnall_2024_excels} and star-forming galaxies at $2 < z < 5$ \citep{mclure_2018_vandels, arellano_cordova_2024_excels_cno, stanton_2024_excels_arh}.
EXCELS-63107 is one of the ancillary targets at $z>6$ selected via strong emission-line signatures in the photometry.
A full description of the target selection and observations for the EXCELS survey is given in \citet{carnall_2024_excels}.

Individual EXCELS targets were observed in subsets of the three NIRSpec gratings.
This strategy was used to maximise the total number of targets observed in the survey \citep[see][]{carnall_2024_excels}.
As far as was possible, individual targets were observed only with gratings that covered relevant rest-frame wavelength intervals.
In the case of EXCELS-63107, which has a redshift of $z=8.271$, only the G395M grating was observed.
At this redshift, G395M covers the key rest-frame optical emission lines that are needed to determine the electron temperature of the gas and the gas-phase oxygen abundance.

The raw level 1 data products were reduced using v1.12.5 of the \emph{JWST} reduction pipeline\footnote{https://github.com/spacetelescope/jwst}.
We ran the default level 1 pipeline with advanced snowball rejection and made use of the CRDS$\_$CTX = jwst$\_$1183.pmap version of the \emph{JWST} Calibration Reference Data System (CRDS) files.
We then ran the level 2 and 3 pipeline steps assuming the default configuration.
The level 3 pipeline combines the 2D spectra from each exposure to produce the final 2D spectrum that can be used for science analysis.
Unfortunately, one of the three dithered exposures for EXCELS-63107 is affected by a shutter that failed to open when commanded, blocking the source.
This failed exposure is not included in the final combination, which means that the total G395M exposure time for EXCELS-63107 is reduced by a factor of $1.5$ to $\simeq 2.7$ hours.
The spectrum was flux calibrated using the \textsc{msafit} tool described in \citet{de_graaff_2024_msafit} assuming a point source morphology as we find that EXCELS-63107 is unresolved in all NIRCam imaging bands (Section~\ref{sec:photometry_and_morphology}).
The final reduced 1D and 2D spectra are shown in Fig.~\ref{fig:spec_and_sed}.

All galaxies in the EXCELS survey benefit from $1$ to $5 \, \mu\mathrm{m}$ NIRCam imaging as part of the PRIMER survey in the F090W, F115W, F150W, F200W, F277W, F356W, F410M and F444W bands \citep{dunlop_2021_primer}.
We have also used the PRIMER \textit{HST} Advanced Camera for Surveys (ACS) imaging mosaics in the F435W, F606W, and F814W bands using data from CANDELS \citep{grogin_2011_candels, koekemoer_2011_candels}.
All bands were point spread function (PSF) homogenised to the F444W imaging.
Full details on the construction of the PRIMER catalogue are given in \citet{carnall_2024_excels}.
Motivated by the point-source morphology of EXCELS-63107 (see Section~\ref{sec:photometry_and_morphology}), we extract photometry using small $0.3$ arcsec-diameter apertures.
To scale to total flux density we then multiply the $0.3$-arcsec photometry by a factor $1.48$ based on a curve-of-growth analysis.

    \begin{figure*}
        \centering
        \includegraphics[width=0.95\linewidth]{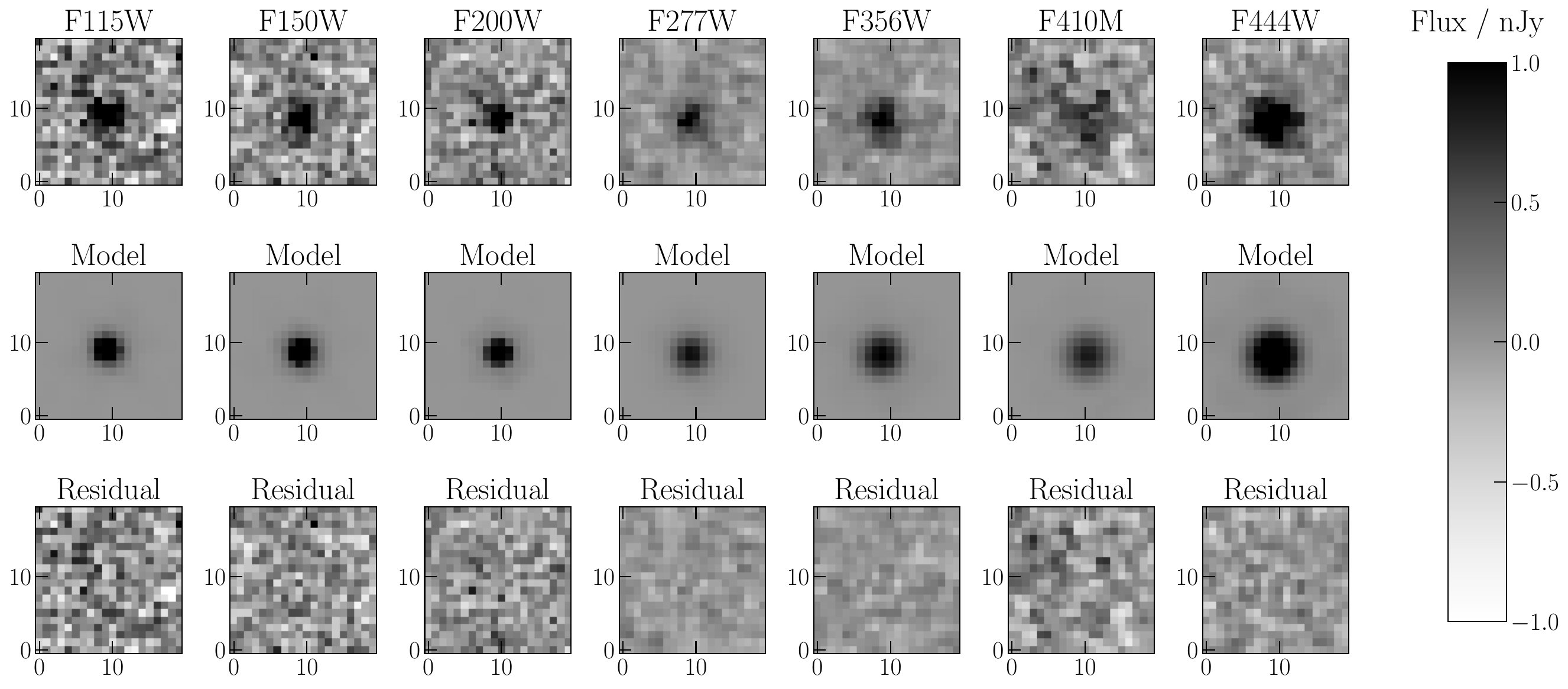}
        \caption{Determining the morphology of EXCELS-63107. An analysis of the \emph{JWST}/NIRCam imaging reveals that EXCELS-63107 is unresolved and can be described by a point-like source in all bands.
        Each column shows one of the NIRCam bands in which EXCELS-63107 is detected (from F115W to F444W) with the rows from top to bottom showing the \emph{JWST}/NIRCam data, the PSF model fit, and the fit residuals respectively.
        The grey scale bar on the far right-hand side shows the flux scale of the images.
        In each band, the PSF fit provides an excellent description of the data from which we conclude that the half-light radius of EXCELS-63107 is $r_e < 200\,\rm{pc}$ (see text).
        The physical size of EXCELS-63107 is therefore comparable with star cluster complexes (or giant \hii \ regions) in the local Universe (e.g. 30 Doradus; $r_e = 100\,\rm{pc}$).
        The lack of any nearby companions suggests that this $z=8.271$ star cluster is forming in isolation.\vspace{0.8cm}} 
        \label{fig:galfit}
    \end{figure*}

\section{The morphology and broadband SED}
\label{sec:photometry_and_morphology}

\subsection{Morphology}
The first notable feature of EXCELS-63107 is its very compact morphology.
To characterise the 2D light distribution we employ the surface-brightness profile fitting software \textsc{galfit} \citep{peng2002,peng2010}. 
We perform fits to all photometry bands in which EXCELS-63107 is detected (i.e. all bands redward of F115W) and assume both a S\'{e}rsic \citep{sersic1968} and PSF (i.e. point-source) profile.
For the S\'{e}rsic fits, we fix the index to $n=1$.
We find that the S\'{e}rsic profile fits provide a generally poor match to the data; with the exception of the F115W band, all return unphysical half-light radii ($r_{e}$) of $\lesssim 0.1 \times$ the full width half maximum (FWHM) of the PSF. 
For the F115W band, the S\'{e}rsic fit yields $r_{e}=0.031\pm0.018^{\prime\prime}$, which is still slightly smaller than the empirical PSF FWHM ($0.037^{\prime\prime}$), but could be considered marginally resolved.
The corresponding physical size is $r_{e}=146\pm85\,\rm{pc}$.
In contrast, all of the PSF fits provide an excellent match to the data (Fig. \ref{fig:galfit}).
For the purposes of this work, we conservatively consider EXCELS-63107 as unresolved in all \emph{JWST}/NIRCam imaging bands, and assume a physical size of $r_{e} < 200\,\rm{pc}$.

The physical extent of EXCELS-63107 is therefore comparable to the sizes of star cluster complexes (also known as giant \hii \ regions) observed in the local Universe, such as 30 Doradus and IIZw40, which have effective radii of ${r_{e} \simeq 100\,\rm{pc}}$ and ${r_{e} \simeq 200\,\rm{pc}}$ respectively \citep{walburn_1991_30dor_size, vanzi_2008_iizw40_size}.
An obvious difference between EXCELS-63107 and these local clusters is its factor $\geq 10 \times$ higher star-formation rate (SFR) and SFR surface density ($\Sigma_{\rm SFR}$).
From the rest-frame optical spectrum (Section~\ref{sec:rest_optical_spectrum}) we infer $\rm{SFR} \simeq 8 \, M_{\odot}yr^{-1}$ (see below) compared to $0.18 \, \rm{M_{\odot}yr^{-1}}$ for 30 Doradus \citep{nayak_2023_30dor_sfr} and $1 \, \rm{M_{\odot}yr^{-1}}$ for IIZw40 \citep{vanzi_2008_iizw40_size}.
The corresponding $\Sigma_{\rm SFR}$ are $\simeq 30, 3$ and $4$ $\rm{M_{\odot}}\rm{yr}^{-1}\rm{kpc}^{-2}$.

\begin{figure}
        \includegraphics[width=0.95\linewidth]{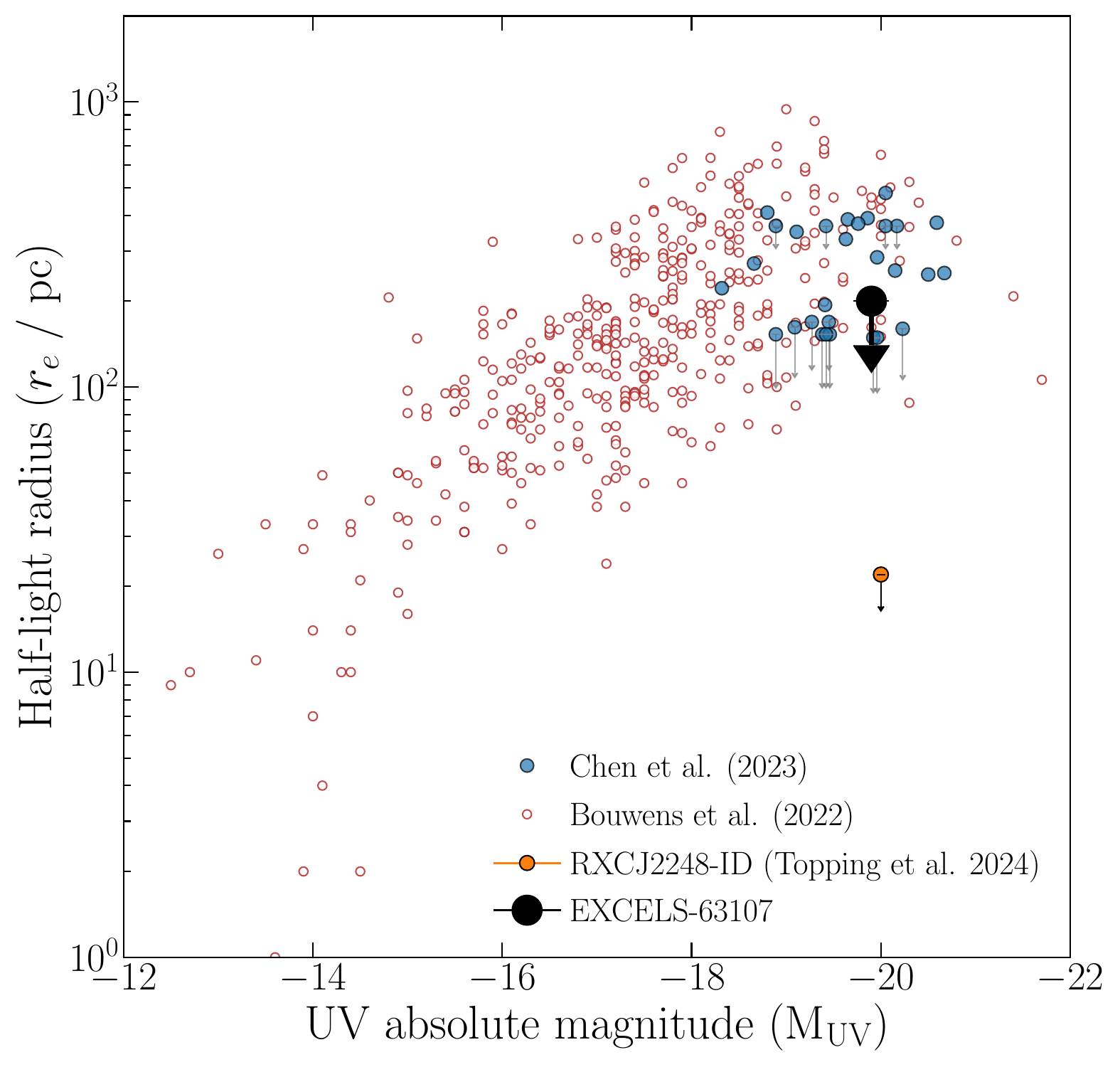}
        \caption{The size and luminosity of EXCELS-63107 compared to other high-redshift sources.
        The black circular data point and arrow show the absolute UV magnitude ($M_{\rm UV} = -19.9$) and upper limit on the half-light radius $r_e < 200 \, \rm{pc}$ that we determine for EXCELS-63107.
        We also include a number of comparison samples from the literature \citep{bouwens_2022_z48_sf_complexes, chen_2023_z68_sf_complexes, topping_2024_dense_z6_galaxies} at $z\simeq4-8$.
        EXCELS-63107 is compact for its UV luminosity, but does not appear to be exceptional compared to other high redshift objects.}
        \label{fig:size_muv}
    \end{figure}

However, EXCELS-63107 does not appear to be exceptional relative to other star-forming complexes seen in galaxies at $z \gtrsim 6$, which commonly have similar, or smaller, sizes and comparable, or larger, $\Sigma_{\rm SFR}$ \citep[e.g.][]{bouwens_2022_z48_sf_complexes, chen_2023_z68_sf_complexes}.
There is certainly evidence for more extreme cases, for example the ${\simeq 20 \, \rm{pc}}$ star-forming clumps observed in the ${z = 6.1}$ lensed galaxy RXCJ2248-ID from \citet{topping_2024_dense_z6_galaxies}, which are estimated to have $\Sigma_{\rm SFR} \sim 10^4 \rm{M_{\odot}}\rm{yr}^{-1}\rm{kpc}^{-2}$ (see also \citealp{vanzella_2023_dense_sf_clumps_sunrise_arc, adamo_2024_z10_dense_star_clusters}).
In Fig.~\ref{fig:size_muv} we show the position of EXCELS-63107 in the $M_{\rm UV} - r_e$ plane, where it can be seen that it falls within a region of the parameter space consistent with previous observations \citep[e.g.][]{bouwens_2022_z48_sf_complexes, chen_2023_z68_sf_complexes}.
Although EXCELS-63107 is arguably one of the most compact objects at its UV luminosity, it may not be an extreme outlier.
Finally, it is worth noting that EXCELS-63107 appears to be a single object without any nearby companions; the closest object with a similar photometric redshift is $\simeq 10 \, \rm{pkpc}$ away.
Barring the existence of an extended low surface-brightness component below the detection threshold (which, in any case, would not be contributing significantly to the UV and optical emission), it appears to be an isolated, compact, star cluster complex.

    \begin{figure}
        \includegraphics[width=0.95\linewidth]{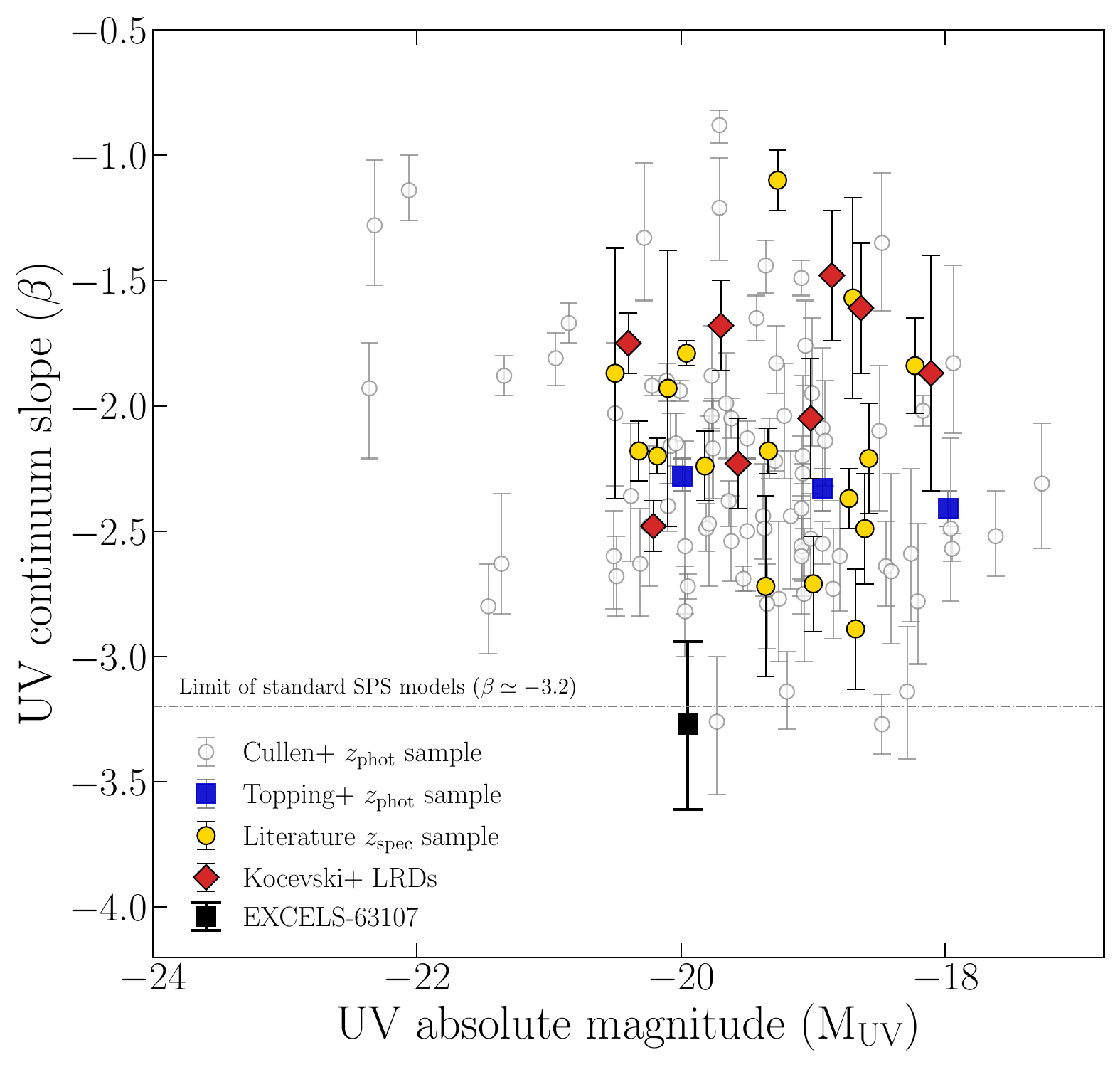}
        \caption{Comparing the UV luminosity $(M_{\mathrm{UV}})$ and continuum slope $(\beta)$ of EXCELS-63107 to literature sources at $z \gtrsim 8$.
        The open grey circular data points are taken from the photometric samples presented in \citet{cullen_2024_uvslope1} and \citet{cullen_2024_uvslope2} (we have restricted their full sample to sources with the best-constrained UV continuum slope measurements [$\sigma_{\beta} \leq 0.3$]).
        The blue square data points show average values of $\beta$ for a photometric sample of $\simeq 200$ galaxies at $8.5 < z_{\rm phot} < 11$ from \citet{topping_2024_uvslopes}. 
        The yellow circular data points show a selection of measurements from the literature in which $\beta$ has been derived from low-resolution NIRSpec/PRISM spectra (extending up to $z\simeq14$; \citealp{curtislake_2023_prismz10, arrabalharo_2023_ceers_z810_prism, carniani_2024_z14}).
        The red diamonds show the LRD sample of \citet{kocevski_2024_lrd} with $M_{\mathrm{UV}} < -18$.
        The black square shows EXCELS-63107, which has $\beta = -3.3 \pm 0.03$ and $M_{\mathrm{UV}} = -19.95$, placing it at the extreme blue end of the $\beta$ distribution, consistent with pure stellar emission and much steeper than expected for AGN models.
        }
        \label{fig:beta_muv}
    \end{figure}

\subsection{UV continuum slope}

The broadband spectral energy distribution (SED) of EXCELS-63107 in the wavelength range ${\lambda_{\rm{obs}} \simeq 0.8 - 5 \, \mu \rm{m}}$ is shown in one of the small inset panels of Fig.~\ref{fig:spec_and_sed}.
An immediately striking feature is the `v-shaped' rest-frame UV to optical photometry, which has become characteristic of compact sources at $z \simeq 4-8$.
These sources have become known as `little red dots' based on their red rest-frame optical colour (LRDs; \citealp{matthee2024_lrds}).
LRDs are most often associated with heavily obscured AGN, which are estimated to constitute $\simeq 70-80 \%$ of the LRD population \citep{labbe_2023_lrds, greene_2024_lrds, kocevski_2024_lrd}.
However, the SED of EXCELS-63107 displays a UV colour that is much steeper than typical AGN-LRDs.
Following the methodology described in \citet{cullen_2024_uvslope2}, we measure a UV continuum slope of $\beta = -3.3 \pm 0.3$.
For comparison, the representative AGN-LRD sample of \citet{kocevski_2024_lrd} has a median UV slope of $\langle \beta \rangle = -1.8$ (see Fig.~\ref{fig:beta_muv}).
The observed UV slope of EXCELS-63107 is in fact more consistent with an extremely young, metal-poor stellar population in the absence of both dust attenuation and strong nebular continuum emission \citep[e.g.][]{cullen_2024_uvslope2}.
The \emph{JWST}/NIRSpec rest-frame optical spectrum (Fig.~\ref{fig:spec_and_sed}) confirms the latter interpretation; rather than being the result of emission from an obscured AGN, the red upturn at optical wavelengths in EXCELS-63107 is caused by strong nebular emission lines emanating from a giant star cluster complex.

The UV spectrum of EXCELS-63107 is not only steep relative to other LRDs, but is also notably steep in comparison to the general population of star-forming galaxies at similar (and even higher) redshifts.
Fig.~\ref{fig:beta_muv} shows the $\beta-M_{\rm UV}$ plane for a selection of $z \gtrsim 8$ galaxy samples from the literature, where it can be seen that EXCELS-63107 sits at the extreme blue end of the $\beta$ distribution, including all spectroscopically confirmed sources up to $z\simeq14$ \citep{carniani_2024_z14} as well as objects up to $\simeq 10 \times$ fainter.
For comparison, the average UV slope of the star-forming galaxy population with $M_{\rm{UV}} \lesssim -18$ is $\langle \beta \rangle \simeq -2.6$ at $z \simeq 10 - 12$ \citep{austin_2024_uvslopes, cullen_2024_uvslope2, morales_2024_uvslopes, topping_2024_uvslopes}.
Interestingly, a UV slope of $\beta = -3.3$ falls just beyond the edge of the parameter space allowed by standard stellar population models, which reach a minimum of $\beta \simeq -3.2$ at the youngest ages and lowest stellar metallicities \citep[e.g.][]{topping_uvslopes_paper1, cullen_2024_uvslope2}.
Accounting for the uncertainty in our measurement, the UV slope of EXCELS-63107 is clearly still fully consistent with these standard models.
However, this relatively extreme UV colour may be the first hint that a unique stellar population is present in this galaxy.

At a minimum, the extremely blue UV slope suggests negligible (essentially zero) dust attenuation as well as a lack of strong nebular continuum emission.
Both dust reddening and the nebular continuum act to flatten (i.e. redden) the UV slope such that, in the presence of either, the observed value is expected to be $\beta \geq -2.6$ \citep[e.g.][]{topping_uvslopes_paper1, cullen_2024_uvslope2, narayanan_2024_dustmodel, katz_2024_balmerjumps}.
The lack of dust attenuation is corroborated by our analysis of the hydrogen Balmer line ratios in the optical spectrum (see Section \ref{subsec:nebular_attenuation}).
It is less straightforward to account for the weak nebular continuum given the fact that there is clearly strong nebular line emission (Fig.~\ref{fig:spec_and_sed}); however, as we show in Section \ref{subsec:cloudy_analysis}, the weak nebular continuum can be explained by invoking a high ionizing photon escape fraction which suppresses the nebular continuum while maintaining the required line luminosities.

\subsection{Stellar mass and star-formation rate}\label{subsec:stellar_mass_estimate}

As we will discuss in detail below (Section \ref{sec:metallicity_and_ionization}), we find that there is convincing evidence that a non-standard stellar population and/or top-heavy IMF is present in EXCELS-63107.
Acknowledging this caveat, we nevertheless estimate the stellar mass and star-formation rate from the broadband SED following a standard approach (i.e. using a standard stellar population model and IMF).
Before fitting the photometry, we first subtract the observed emission-line fluxes measured from the spectrum (see Section \ref{sec:rest_optical_spectrum} for a description of the emission-line fitting).
Unfortunately, we cannot correct for any rest-frame UV emission lines that may be present, and therefore our SED modelling explicitly assumes that any UV lines do not strongly affect the UV photometry (this should be a reasonable approximation unless strong \lya \ emission is present)\footnote{We note that the SED fitting results are unchanged if we exclude the F115W filter which covers the \lya \ region.}.

We fit the resulting stellar SED using \textsc{bagpipes} \citep{carnall_2018_bagpipes} assuming the \citet{bc03_stellar_models} stellar population models, a \citet{chabrier_imf} IMF and metallicities between $Z=0.007\,\mathrm{Z}_{\odot}$ and $Z=2.5 \, \mathrm{Z}_{\odot}$.
We fix the dust attenuation to zero based on the blue UV continuum slope and the observed Balmer line ratios (see below).
We test two star-formation histories (SFHs) in order to assess the full range of stellar masses consistent with the photometry.
First, we assume a standard double power law SFH to model a scenario in which stellar mass is built up steadily; this assumption yields a stellar mass of $\mathrm{log}(M/\mathrm{M}_{\odot})=8.66^{+\,0.13}_{-0.17}$ and a star-formation rate of $2.2^{+\,0.3}_{-0.2} \, \mathrm{M_{\odot}yr}^{-1}$ (averaged over the past 10 Myr).
Second, we add a young 3 Myr burst to the SFH model to approximate a scenario in which the UV emission is dominated by a recently formed population.
Under the burst model assumption, we derive a similar total stellar mass of $\mathrm{log}(M/\mathrm{M}_{\odot})=8.57^{+\,0.32}_{-1.03}$ and a star-formation rate of $2.8^{+\,0.7}_{-0.7} \, \mathrm{M_{\odot}yr}^{-1}$.
In this burst model, the stellar mass formed in the 3 Myr burst is $\mathrm{log}(M_{\mathrm{burst}}/\mathrm{M}_{\odot})=7.35^{+\,0.22}_{-1.30}$ (i.e., roughly $10$ per cent of the total stellar mass).

The addition of the young burst allows the stellar mass to be much lower, and this large uncertainty highlights the well-known difficulties associated with estimating stellar masses for star-forming objects using only rest-frame UV and optical data\footnote{Unfortunately EXCELS-63107 falls outside of the PRIMER UDS region with overlapping MIRI $\lambda = 7.7 \, \mu \mathrm{m}$ photometry; the longest wavelength probed is $\lambda \simeq 4.4 \, \mu \mathrm{m}$ (corresponding to $\lambda_{\rm rest} \simeq 0.55 \, \mu \mathrm{m}$).}.
Based on the compact morphology of EXCELS-63107, its ultra-blue UV continuum, and extremely low metallicity (see Section \ref{sec:metallicity_and_ionization}), we expect that the extended star-formation plus recent burst model is likely to be a more accurate description of the physical reality.
Recognizing again the significant systematic uncertainties at play, we nevertheless adopt a total stellar mass of $\mathrm{log}(M/\mathrm{M}_{\odot})=8.57^{+\,0.32}_{-1.03}$ with a burst mass of ${\mathrm{log}(M_{\mathrm{burst}}/\mathrm{M}_{\odot})=7.35^{+\,0.22}_{-1.30}}$ as our best estimate.

\section{The rest-frame optical spectrum}
\label{sec:rest_optical_spectrum}

The NIRSpec/G395M spectrum of EXCELS-63107 is shown in Fig. \ref{fig:spec_and_sed}. 
This spectrum covers the rest-frame wavelength range ${\lambda_{\rm rest} = 3100-5500\,}${\AA} enabling measurements of the key emission lines from \oii \ up to \oiiia.
To measure the emission-line fluxes we follow the methodology outlined in detail in a companion paper \citep{scholte_2025_excels}.
Briefly, we estimate and subtract the continuum flux using the running mean of the $16^{\rm th}$ to $84^{\rm th}$ percentile flux values within a rest-frame wavelength window with $\Delta \lambda = 350 \,${\AA}. 
From the resulting continuum-subtracted spectrum, all of the available emission lines are fit simultaneously; we allow the line amplitude to vary and assume a common intrinsic line width and line velocity.
The best fitting line fluxes are determined via a $\chi^2$-minimization and the uncertainties are estimated using a line profile weighted average of the individual pixel flux uncertainties \citep{moustakas_2023_linefit_desi}.
The \oii \ line is not detected (Fig.~\ref{fig:spec_and_sed}); in this case, we estimate a $2\sigma$ upper limit by inserting a line of increasing flux into the continuum-subtracted spectrum (at the correct wavelength) until a $2\sigma$ detection is achieved.
The observed line fluxes, and the upper limit on \oii, are given in Table~\ref{table:line_fluxes}.

Before discussing this spectrum in more detail, it is worth highlighting some important features that are immediately visible in Fig.~\ref{fig:spec_and_sed} (and evident from the values in Table~\ref{table:line_fluxes}).
First, the \oiiia/\oii \ ratio is large, indicative of a highly-ionized \hii \ region.
We measure a $2\sigma$ lower limit of \oiii/\oiinwl$\, >  15$.
Combined with the compact morphology, this ratio is suggestive of a small volume of highly ionized gas with no neutral outer boundary (i.e. a density bounded \hii \ region; \citealp{nakajima_2013_ism_ionisation}).
Second, the auroral \oiiiaur \ line is robustly detected ($\rm{S/N}=4.7$) and is stronger than is typical for other star-forming galaxies.
We measure an auroral to forbidden line ratio of \oiiiaur/\oiiia$\,=0.074 \pm 0.016$ which, to our knowledge, is the largest such ratio observed in any galaxy to date (typical values are $\lesssim 0.05$).
This ratio is sensitive to the electron temperature and density of the ionized gas and the value we measure suggests an extremely hot, or extremely dense, \hii \ region  \citep[e.g.][]{katz_2023_strongo3aur}\footnote{The ratio becomes density-sensitive at $n_e > 10^4 \, \rm{cm}^{-3}$.}.
Third, the \oiiia/\hbeta \ ratio of $3.61 \pm 0.31$ is relatively modest in comparison to typical values observed in high-redshift galaxies (at $z\simeq2-3$ the typical \oiiia/\hbeta \ ratio for star-forming galaxies with $\mathrm{log}(M_{\star}/\mathrm{M}_{\odot}) = 9$ is $\simeq 5-10$; e.g. \citealp{cullen_2021_alpha_enhancement, sanders_2021_mosdef_mzr}).
This weak $\rm{O}^{2+}/\rm{H}$ ratio, combined with the lack of $\rm{O}^{+}$ in the spectrum, suggests a low overall gas-phase oxygen abundance.
A low metallicity would also be consistent with the hot gas temperature implied by the \oiiiaur/\oiiia \ ratio.
Together, these features suggest a highly-ionized, low-metallicity \hii \ region that is either extremely hot or extremely dense.
In Section~\ref{sec:metallicity_and_ionization}, we perform a detailed modelling analysis to determine the most likely physical interpretation of this unusual spectrum.

    \begin{table}
        \centering
        \caption{Observed line fluxes for EXCELS-63107.}
         \label{table:line_fluxes}
         \renewcommand{\arraystretch}{1.25}
        \begin{tabular}{lr}
            \hline
            \hline
            Line & Observed Flux / $10^{-19}$ erg s$^{-1}$ cm$^{-2}$ \\
            \hline
            \oii & $< 1.47^a$ \\
            \neiiia & $3.11 \pm 0.59$ \\
            \neiiib$^b$ & $2.48 \pm 0.58$ \\
            \hdelta & $2.17 \pm 0.55$ \\
            \hgamma & $6.62 \pm 0.80$ \\
            \oiiiaur & $2.84 \pm 0.61$ \\
            \hbeta & $10.69 \pm 0.84$ \\
            \oiiib & $13.81 \pm 0.92$ \\
            \oiiia & $38.54 \pm 1.20$ \\
            \hline
        \multicolumn{2}{l}{$^a$ $2\sigma$ upper limit.}\\
        \multicolumn{2}{l}{$^b$ Contaminated by the \heps \ line.}\\
        \end{tabular}
    \end{table}

\subsection{Nebular dust attenuation and star-formation rate}\label{subsec:nebular_attenuation}

The hydrogen Balmer series lines \hbeta, \hgamma \ and \hdelta \ are detected with $\rm{S/N} > 3$ in the spectrum and can be used to assess the level of nebular dust attenuation.
We measure Balmer line ratios of \hgamma/\hbeta$\,= 0.62 \pm 0.09$ and \hdelta/\hbeta$\,=0.20 \pm 0.05$ both of which are within $\simeq 1.5\sigma$ of their theoretical Case B recombination ratios assuming typical values of $n_e=100\,\mathrm{cm}^{-3}$ for the electron density and $T_e=10 \, 000 \, \rm{K}$ for the electron temperature (i.e. \hgamma/\hbeta$\,=0.47$ and \hdelta/\hbeta$\,=0.26$).
As we discuss in Section \ref{sec:metallicity_and_ionization}, our photoionization models (which also assume no dust) predict a much higher gas temperature ($T_e \simeq 35 \, 000 \, \rm{K}$) but yield only slightly different intrinsic ratios (\hgamma/\hbeta$\,=0.48$ and \hdelta/\hbeta$\,=0.27$).
Based on these Balmer line ratios and the ultra-blue UV slope, it is reasonable to conclude that both the stellar and nebular emission are essentially unattenuated by dust.

Assuming no nebular attenuation, we use the observed \hbeta \ flux to derive a star-formation rate of $7.8 \pm 0.6 \, \mathrm{M_{\odot}yr}^{-1}$.
Here, we use the intrinsic \halpha/\hbeta \ ratio of our best-fitting photoionization model (Section \ref{sec:metallicity_and_ionization};  \halpha/\hbeta$\,=2.68$) to convert to \halpha \ flux, and use the \halpha \ luminosity to convert to a star-formation rate applying a conversion appropriate for low-metallicity stellar populations from \citet{reddy_2018_low_metal_sfr_conversion}\footnote{$\mathrm{SFR(H\alpha)}=3.236 \times 10^{-42} \, \mathrm{L(H\alpha)}$.}.
This SFR estimate is larger than the one derived from the SED modelling, but some difference is perhaps unsurprising given the systematic uncertainties involved in the SED fitting.  
Combined with our best estimate of the stellar mass, the \hbeta-based star-formation rate converts into a specific star-formation rate (sSFR) of $\simeq 20 \, \mathrm{Gyr}^{-1}$, implying a mass doubling time of $\simeq 50 \, \mathrm{Myr}$; these values are roughly consistent with the average value for galaxies at the same redshift \citep[e.g.][]{topping_2022_z78_ssfr}.

    \begin{figure}
        \includegraphics[width=0.95\linewidth]{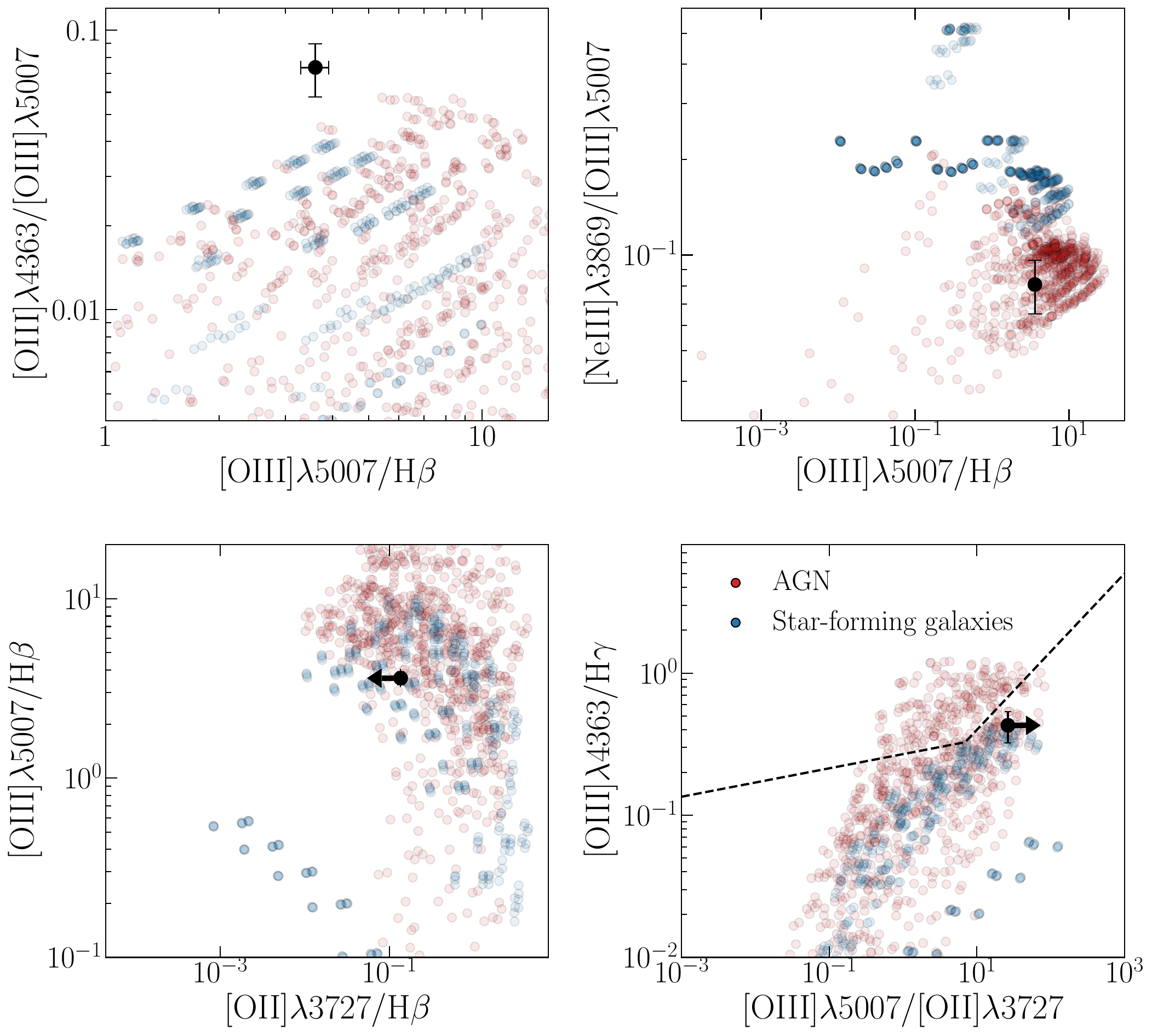}
        \caption{A comparison between the observed line ratios in EXCELS-63107 and the photoionization models of \citet{nakajima_2022_photoion_models}.
        Each panel shows a different emission-line ratio diagnostic diagram with the EXCELS measurements shown in black.
        For diagrams involving the \oii \ line an arrow is shown to represent the $2 \sigma$ limit.
        The blue and red points show the photoionization models of \citet{nakajima_2022_photoion_models} representing the predictions of star-forming galaxies and AGN, respectively.
        In most cases, the results are ambiguous, with both star-formation and AGN models consistent with the observations.
        The one exception is the \neiiia/\oiiia \ versus \oiiia/\hbeta \ diagram (upper-right panel); however, as discussed in Section \ref{subsec:cloudy_analysis}, our photoionization models demonstrate that stellar models can account for the observed \neiiia/\oiiia \ ratio.
        The dashed line in the bottom right panel is the demarcation line between AGN and star-formation derived by \citet{mazzolari_2024_o3aur_hg_agn_diagnostic}, where EXCELS-63107 falls within the star-forming region.
        }
        \label{fig:nakajima_models}
    \end{figure}
    
\subsection{Is EXCELS-63107 an AGN?}

Before continuing, it is worth considering whether it is likely that EXCELS-63107 hosts an AGN.
Deep \emph{JWST} spectroscopy of $z>6$ galaxies has revealed evidence for active galactic nuclei (AGN) with black hole (BH) masses as low as $\mathrm{log}(M_{\mathrm{BH}}/\mathrm{M}_{\odot}) \simeq 5.6$ \citep[e.g.][]{harikane_2023_lowmass_agn, maiolino_2023_jades_agn, greene_2024_lrds, matthee2024_lrds}.
At the low metallicities expected at these redshifts ${(Z/\rm{Z}_{\odot} \lesssim 0.2)}$, it is difficult to distinguish these AGN from regular star formation using standard line ratio diagnostics \citep[e.g.][]{ubler_2023_agn_z5p55}.  
Moreover, they are often undetected at X-ray wavelengths, making them especially challenging to identify \citep[e.g.][]{maiolino_2024_agn_xray_nondetections}.  

Taking these considerations into account, it may ultimately be impossible to definitively rule out the presence of a low-mass AGN in EXCELS-63107.
However, it is also true that the data do not display strong evidence in favour of an AGN.
The ultra-steep UV slope ($\beta=-3.3 \pm 0.3$) is probably the clearest sign that an AGN is not strongly contaminating the spectrum.
UV slopes this steep are inconsistent with emission from AGN accretion disks, which are expected to emit a UV spectrum with $\beta \geq -2.4$ (e.g. \citealp{shakura_1973_agn_seds, cheng_2019_agn_seds}; a standard, dust-free, Shakura $\&$ Sunyaev accretion disk has $\beta = -7/3 \simeq -2.33$).
We also do not see any evidence of broadening in the permitted \hbeta, \hgamma \ or \hdelta \ emission lines, suggesting that an AGN, if present, must be very low mass and is not likely to dominate the UV and optical emission.
Finally, a comparison to the \citet{nakajima_2022_photoion_models} photoionization models across four line ratio diagnostic diagrams yields mixed results (Fig. \ref{fig:nakajima_models}).
In most cases, the line ratios are ambiguous and can be explained by star formation or AGN.
The one exception is the \neiiia/\oiiia \ versus \oiiia/\hbeta \ diagram, in which the data fall within the pure-AGN region, albeit still within $< 2 \sigma$ of star-forming models.
However, as we discuss in Section \ref{sec:metallicity_and_ionization}, our bespoke photoionization models reproduce the \neiiia/\oiiia \ ratio without invoking AGN emission.
It is also worth noting that EXCELS-63107 falls outside the pure AGN region of the \oiiiaur/\hgamma \ versus \oiiia/\oii \ diagram based on the demarcation line recently proposed by \citet{mazzolari_2024_o3aur_hg_agn_diagnostic}.
Overall, we conclude that there is no strong evidence to suggest that EXCELS-63107 hosts an AGN.
In what follows, we assume that the emission comes from a star cluster complex powered by massive stars.

\section{Gas-phase metallicity and the ionizing source}
\label{sec:metallicity_and_ionization}

\subsection{The gas-phase oxygen abundance}
\label{subsec:pyneb_analysis}

The detection of the \oiiiaur \ auroral line allows us to directly estimate the electron temperature ($T_e$) of the gas within the \hii \ region.
With a robust estimate of $T_e$, it is then possible to derive accurate gas-phase element abundances.

Typically, \hii \ regions are modelled assuming a three-zone ionization structure composed of low-, intermediate-, and high-ionization zones.
In extreme cases, a very high-ionization zone is also considered \citep[e.g.][]{berg_2021_four_zone_ionisation_model}.
The metal lines detected in the spectrum of EXCELS-63107 correspond to the ions $\rm{O}^{2+}$ and $\rm{Ne}^{2+}$, which are associated with the high-ionization zone (ionization potential energies of $\simeq 35-60\, \rm{eV}$; \citealp{berg_2021_four_zone_ionisation_model}).
As a result, our data are only sensitive to the temperature of the gas in this zone. 
However, the fact that we do not detect the \oii \ line (i.e. $\rm{O}^{+}$) implies that the volume of the low- and intermediate-ionization zones is likely to be negligible, and that the majority of the gas is in the highly ionized state.
Our photoionization modelling analysis generally confirms this ionization structure (see Section \ref{subsec:cloudy_analysis}).
As a result, the ionic abundances of the high-ionization species therefore turn out to be good approximations of the total element abundances (i.e. $\rm{O}^{2+}/H^+ \simeq \rm{O}/H$; although see Section \ref{subsec:cloudy_analysis} for a more detailed discussion).

Once the temperature is known, the abundances of a given ion relative to hydrogen can be determined using:
\begin{equation}
    \label{eq:ionic_abundances}
    \frac{N(X^i)}{N(\mathrm{H^+})}=\frac{f_{\lambda, i}}{f_{H\beta}}\frac{j_{\mathrm{H}\beta}(T_e, n_e)}{j_{\lambda, i}(T_e, n_e)}
\end{equation}
where $f_{\lambda, i}$ is the flux of an emission line associated with the ionization state $i$ (e.g. \oiiia \ in the case of $\rm{O}^{2+}$) and $j_{\lambda, i}$ is the line emissivity, which is a function of both the electron temperature and density ($n_e$).
Unfortunately, there are no direct density-sensitive features in the spectrum of EXCELS-63107; however, we overcome this by adopting a forward modeling approach that allows us to marginalise over a range of densities.

    \begin{table}
        \centering
        \caption{The uniform priors assumed in our determination of the oxygen and neon abundances of EXCELS-63107.}
         \label{table:priors}
         \renewcommand{\arraystretch}{1.25}
         \begin{tabular}{lll} 
            \hline
            \hline
            Parameter & Description & Prior \\
            \hline
            $\rm{log(O^{2+}/H^+)}$ & abundance of $\rm{O^{2+}}$ relative to H & $\mathcal{U}(-7, -2)$ \\
            $\rm{log(Ne^{2+}/H^+)}$ & abundance of $\rm{Ne^{2+}}$ relative to H & $\mathcal{U}(-7, -2)$ \\
            $\mathrm{log}(T_e)$ & high-ionization zone temperature & $\mathcal{U}(3, 5)$ \\
            $\mathrm{log}(n_e)$ & electron density & $\mathcal{U}(1, 6)$ \\
            \hline
        \end{tabular}
    \end{table}

We model the line emissivities using the emission-line analysis software \textsc{pyneb} \citep{luridiana_2015_pyneb}.
For the $\rm{O}^{2+}$ ion, we adopt the transition probabilities from \citet{froese_fischer_2004_o2p_transitions} and collision strengths from \citet{aggarwal_1999_o2p_collisions}.
For $\rm{Ne}^{2+}$, the transition probabilities and collision strengths are both taken from \citet{mcLaughlin_2011_ne2p_trans_collisions}.
For \hbeta, we use the \textsc{pyneb} core function \texttt{getHbEmissivity} which is based on the formula given in \citet{aller_1984_hbemiss_reference}.
Our model comprises four free parameters: $\rm{log(O^{2+}/H^+)}$, $\rm{log(Ne^{2+}/H^+)}$, $\mathrm{log}(T_e)$ and $\mathrm{log}(n_e)$, where $\rm{O^{2+}/H^+}$ and $\rm{Ne^{2+}/H^+}$ are the ionic abundances relative to hydrogen (i.e. the left-hand term in equation~\ref{eq:ionic_abundances}).
We assume no dust attenuation (see Section~\ref{sec:rest_optical_spectrum}).
Based on these four parameters, we can predict the \oiiiaur/\oiiia, \oiiia/\hbeta \ and \neiiia/\hbeta \ flux ratios (equation~\ref{eq:ionic_abundances}) and compare directly to the observed values.
We use the nested sampling code \textsc{dynesty} \citep{speagle_2022_dynesty, koposov_dynesty_zenodo} to sample the parameter space and determine the posterior probability distributions, adopting a standard Gaussian likelihood.
We assume flat priors on all parameters as outlined in Table \ref{table:priors}.

\begin{figure}
        \includegraphics[width=0.95\linewidth]{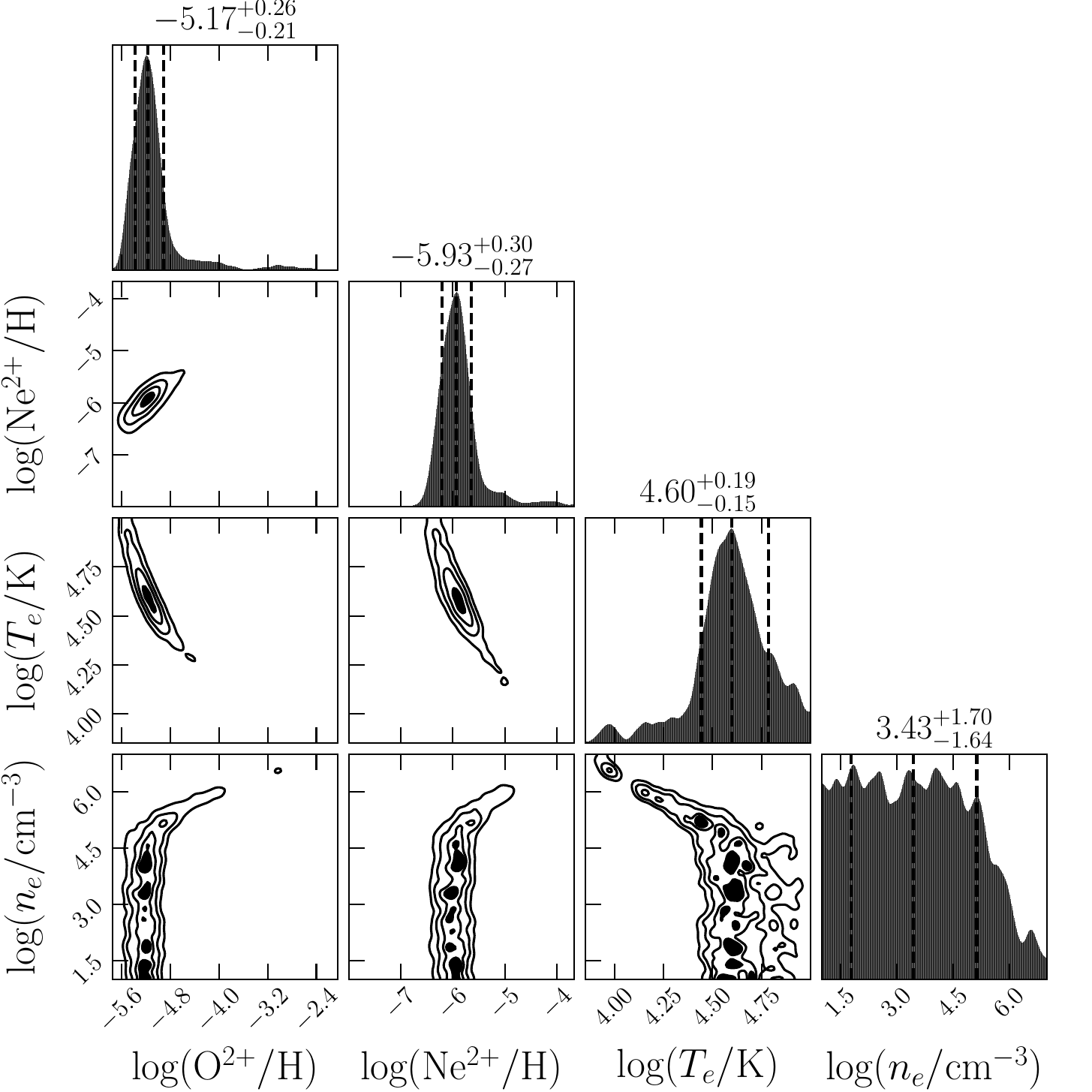}
        \caption{Corner plot showing the 1D and 2D marginalised posterior distributions for our 4-parameter \textsc{pyneb} model.
        The 16th, 50th and 84th percentiles (i.e. the $1\sigma$ credible intervals) are shown by the vertical dashed lines and the corresponding parameter constraints are given above the 1D posterior panels.
        Our data can constrain the electron temperature ($T_e$) and the abundances of $\rm{O}^{2+}$ and $\rm{Ne}^{2+}$.
        The constraint on the electron density ($n_e$) is much weaker, but extremely large densities of $n_e \gtrsim 10^5\,\rm{cm}^{-3}$ are disfavoured.
        }
        \label{fig:dynesty_corner}
    \end{figure}

The resulting posterior distributions for the four model parameters are shown in Fig. \ref{fig:dynesty_corner} where it can be seen that the abundances of $\rm{O^{2+}/H}$, $\rm{Ne^{2+}/H^+}$, and $T_e$ are well constrained by the data.
Focusing first on the $\rm{O}^{2+}$ abundance, we find $\rm{log(O^{2+}/H^+)}=-5.17^{+0.26}_{-0.21}$.
To relate this ionic abundance to the total oxygen abundance, we assume $\rm{O}^{+}/H^+ \simeq 0$ (based on the non-detection of \oii) and also assume that the $\rm{O}^{3+}/H^+$ contribution is negligible.
The first of these assumptions is validated by our photoionization modelling, but see Section \ref{subsec:cloudy_analysis} for a discussion of the possible contribution of higher ionization states.
For now, we assume that the doubly-ionized oxygen abundance is a good approximation to the total oxygen abundance (i.e. $\rm{O}^{2+}/H \simeq \rm{O}/H$)\footnote{Including the \oii \ upper limit in this analysis yields an upper limit on the $\rm{O}^+$ abundance of $\rm{log(O^{+}/H^+)}= < -6.7$. The data therefore indicate that the $\rm{O}^+$ abundance is at least an order of magnitude lower than the $\rm{O}^{2+}$ abundance.
Our subsequent photoionization model analysis suggest that the fraction of $\rm{O}^+$ in the \hii \ region is approximately zero (Section \ref{subsec:cloudy_analysis})}.

Converting to more familiar units, we find ${12+\rm{log(O/H)} = 6.83^{+0.26}_{-0.21}}$.
Taking the abundance of oxygen as a proxy for the total metallicity, we find ${Z = 0.0139^{+0.011}_{-0.005}\, \rm{Z_{\odot}}}$ (i.e, ${\simeq 1.4 \%}$ of the solar value \citealp{asplund_2021_solar_abn}).
EXCELS-63107 therefore falls firmly into the category of an `Extremely Metal-Poor Galaxy' (EMPG), which are typically defined as galaxies with $Z < 0.1\,\rm{Z}_{\odot}$ (the $95\%$ confidence interval for the metallicity of EXCELS-63107 is between $0.5 - 3\%$ solar).
In fact, the best-fitting value is one of the lowest direct-method metallicities measured in a galaxy to date\footnote{We note that it is very plausible that the stellar complex reported in \citet{vanzella_pop3_like_cluster} has a lower metallicity than EXCELS-63107, but their inference of ${12+\rm{log(O/H)} < 6}$ is based on strong line diagnostics. In this paper, we primarily focus our comparison on other galaxies with direct temperature-based metallicity estimates.}.
Before this, the lowest direct metallicity measured was ${12+\rm{log(O/H)} = 6.90\pm0.03}$ (i.e. ${Z \simeq 0.016 \, \rm{Z_{\odot}}}$) in a dwarf galaxy at ${z=0.03}$ (\citealp{kojima_2020_empgs}; although also see \citealp{thuan_2022_revision_of_kojima} who subsequently revised the abundance upwards to $12+\mathrm{log(O/H)} = 7.13\pm0.03$).
In Section \ref{sec:discussion_of_excels63107}, we provide a more detailed comparison between EXCELS-63107 and these local Universe EMPGs, as well as discussing EXCELS-63107 in the context of other $z\simeq7-9$ galaxies.

The $\rm{Ne}^{2+}$  abundance is also well-constrained by our data to be $\rm{log(Ne^{2+}/H^+)}=-5.93^{+0.30}_{-0.27}$.
As with oxygen, the conversion from ionic to total abundance is also relatively simple.
For a highly-ionized \hii \ region in which the majority of oxygen is in the $\rm{O}^{2+}$ state, the majority of neon will be in the $\rm{Ne}^{2+}$ state \citep[e.g.][]{izotov_2006_abundances_and_icfs, dors_2013_neon_icf} so that ${\rm{Ne}^{2+}/H \simeq \rm{Ne}/H}$.
Our estimate for the neon abundance of EXCELS-63107 is therefore ${12+\rm{log(Ne/H)} = 6.08^{+0.30}_{-0.27}}$, which is also $\simeq 1 \%$ of the solar value \citep{asplund_2021_solar_abn}.
The resulting neon-to-oxygen abundance ratio is $\rm{log(Ne/O)}=-0.79^{+0.11}_{-0.14}$, within $2\sigma$ of the \citet{asplund_2021_solar_abn} solar ratio ({$\rm{log(Ne/O)}_{\odot}=-0.63$}).
Our analysis is therefore consistent with other literature studies that find solar-like Ne/O abundances in high-redshift galaxies \citep[e.g.][]{arellano-cordova-2024-ers-abundances-z78, stanton_2024_excels_arh}, as expected for the pure $\alpha$-elements \citep{kobayashi_2020_gce_model}.

    \begin{table}
        \centering
        \caption{Best-fitting physical parameters for EXCELS-63107.} 
        \label{table:physical_params}
        \renewcommand{\arraystretch}{1.25}
        \begin{tabular}{lr} 
        \hline
        \hline
        Parameter & Value \\
        \hline
        $M_{\rm UV}$ & $-19.9 \pm 0.1$ \\
        $r_e / \, \rm{pc}$ & $< 200$ \\
        $\mathrm{log}(M/\mathrm{M_{\odot}})$ (double power law [DPL]) & $8.66^{+\,0.13}_{-0.17}$ \\
        $\mathrm{log}(M/\mathrm{M_{\odot}})$ (DPL + burst model) & $8.57^{+\,0.32}_{-1.03}$ \\
        $\mathrm{log}(M_{\mathrm{burst}}/\mathrm{M_{\odot}})^{\rm a}$ & $7.35^{+\,0.22}_{-1.30}$ \\
        $\mathrm{SFR}_{\mathrm{H}\beta} / \, \mathrm{M_{\odot}yr}^{-1}$ & $7.8 \pm 0.6$ \\
        \hline
        \multicolumn{2}{c}{\textsc{pyneb} forward modelling results}\\
        \hline
        $\mathrm{log(O^{+}/H^+)}$ & $< 6.7 \, (2\sigma)$ \\
        $\mathrm{log(O^{2+}/H^+)}$ & $-5.17^{+0.26}_{-0.21}$ \\
        $12+\mathrm{log(O/H)}$ & $6.83^{+0.26}_{-0.21}$ \\
        $12+\mathrm{log(O/H)}$ (ICF correction) & $6.89^{+0.26}_{-0.21}$ \\
        $\mathrm{log(Ne^{2+}/H^+)}$ & $-5.93^{+0.30}_{-0.27}$ \\
        $\mathrm{log(Ne/O)}$ & $-0.79^{+0.11}_{-0.14}$ \\
        $\mathrm{log}(n_e/\mathrm{cm}^{-3})$ & $3.5^{+2.1}_{-1.2}$ \\
        $T_e / 10^4 \, \rm{K}$ & $3.9^{+2.1}_{-1.2}$ \\
        \hline
        \multicolumn{2}{l}{$^{\rm a}$ The estimated stellar mass of the burst in the DPL + burst model.}\\
        \end{tabular}
    \end{table}

Perhaps the most surprising result from Fig. \ref{fig:dynesty_corner} is the extremely high electron temperature inferred for the ionized gas.
We find ${\mathrm{log}(T_e / \mathrm{K}) = 4.60^{+0.19}_{-0.15}}$ or ${T_e = 3.9^{+2.1}_{-1.2}\times10^4 \, \rm{K}}$ (Fig. \ref{fig:dynesty_corner}).
If this is true, it implies that \emph{the electron temperature within the ionized volume is $\simeq 40, 000 \, \rm{K}$, as hot as the surface temperature of a typical O-type star.}
A corollary of this is that the ionizing source must be much hotter than a typical O star.
This extremely high inferred temperature is fundamentally driven by the large \oiiiaur/\oiiia \, ratio observed in the spectrum.
However, a high \oiiiaur/\oiiia \, ratio can be caused by extremely hot temperatures \emph{or} high densities ($\gtrsim 10^5 \, \mathrm{cm}^{-3}$; \citealp{katz_2023_strongo3aur}).
As can be seen from Fig. \ref{fig:dynesty_corner}, our data favour the high-temperature interpretation.
The primary reason for this is that a \hii \ region with $n_e \gtrsim 10^5~\mathrm{cm}^{-3}$ should exhibit boosted \neiiia \ (critical density $1 \times 10^7~\mathrm{cm}^{-3}$) relative to \oiiia \ (critical density $7 \times 10^5~\mathrm{cm}^{-3}$) as a result of increasing collisional de-excitation of the $2p$ state in $\rm O^{2+}$.
That being said, the posterior probability distributions in Fig. \ref{fig:dynesty_corner} do not completely rule out the high density, low temperature interpretation.
However, as we will describe in Section \ref{subsec:breaking_te_ne_degeneracy}, we can appeal to the extremely blue UV continuum slope to help to distinguish between these scenarios. 

In summary, the \textsc{pyneb} analysis suggests an extreme physical scenario in which an exotic ionizing source is heating low metallicity gas ($\simeq 1.5\%$ solar) to ${T_e \simeq 40,000 \, \rm{K}}$.
All of the key derived parameters are given in Table \ref{table:physical_params}.
Below, we perform a detailed photoionization modelling analysis to determine whether such a scenario is physically plausible.

\subsection{\textsc{cloudy} photoionization models}
\label{subsec:cloudy_analysis}

    \begin{table}
        \centering
        \caption{Parameter ranges/values used for the \textsc{cloudy} photoionization modeling.} 
        \label{table:cloudy}
        \renewcommand{\arraystretch}{1.25}
        \begin{tabular}{lrr} 
        \hline
        \hline
        Parameter & Range or Value(s)  & Step size \\
        \hline
        $r_{\rm{HII}}^{\rm a}$ & $[20, 50, 100, 150] \, \rm{pc}$ & $--$ \\
        $r_{\rm{inner}}^{\rm b}$ & $3 \, \rm{pc}$ & $--$ \\
        $\mathrm{log}(Z/\rm{Z}_{\odot})$ & $-2.1 \leq \mathrm{log}(Z/\rm{Z}_{\odot}) \leq -0.7$ & $0.1$ \\
        $\mathrm{log}(n_e / \rm{cm}^{-3})$ & $2 \leq \mathrm{log}(n_e / \rm{cm}^{-3}) \leq 6$ & $1.0$ \\
        $\mathrm{log(Ne/O)}$ & $-0.79$ & $--$ \\
        $\mathrm{log(O/Fe)}$ & $0.48$ & $--$ \\
        \hline
        \multicolumn{3}{l}{$^{\rm a}$ The maximum outer radius of the \hii \ region.}\\
        \multicolumn{3}{l}{$^{\rm b}$ The inner radius of the \hii \ region (fixed).}\\
        \end{tabular}
    \end{table}

Although an emission-line analysis code like \textsc{pyneb} can make predictions for temperatures and densities based on observed line ratios and emissivities, it cannot self-consistently account for the ionization structure and cooling within realistic \hii \ regions.
Fortunately, because of the well-understood properties of EXCELS-63107, it is possible to produce a relatively realistic physical model. 
Specifically, we know that the galaxy is extremely compact ($\lesssim 200\,\rm{pc}$) and likely to be a single, young star-forming complex (Section \ref{sec:photometry_and_morphology}); we also know that we can reasonably ignore the effects of dust (Section \ref{sec:rest_optical_spectrum}).

We use the photoionization code \textsc{cloudy} (version 23.01; \citealp{cloudy_2023_version}) to model a range of spherical \hii \ regions with radii between $20-150\,\rm{pc}$, gas-phase metallicities between $0.5 - 10\%$ solar, and electron densities between $10^2-10^6\,\rm{cm}^{-3}$.
For all models, we adopt an inner radius of $3 \, \rm{pc}$.
The calculation stops when the outer radius is reached ($r_{\rm{HII}}^a$ in Table \ref{table:cloudy}) or when the electron fraction drops below $0.01$ (i.e. when the edge of the \hii \ region is formed within the maximum allowed radius).
These two distinct stopping criteria are crucial as they allow us to model both density-bounded and ionisation-bounded \hii \ regions.

We assume a solar-like chemical composition for the gas but change the ratios of nitrogen and carbon relative to oxygen to reflect the typical values observed in $z>6$ systems ($\mathrm{log(C/O)} \simeq -1.0$ and $\mathrm{log(N/O)} \simeq -0.5$; \citealp{arellano-cordova-2024-ers-abundances-z78, jones_2023_co_abundance, topping_2024_dense_z6_galaxies, curti_2024_z9_cno, marques-chaves-2024-highz-n-emitters}).
We also scale the iron abundances such that $\mathrm{O/Fe} = 2.5 \times (\mathrm{O/Fe})_{\odot}$ to simulate the $\alpha$-enhanced abundances known to be ubiquitous, and expected, at high redshifts \citep[e.g.][]{steidel_2016_fuv_optical, cullen_2019_feh_mzr, topping_2020_alpha_enhancement, cullen_2021_alpha_enhancement, kashino_2022_stellar_mzr, strom_2022_ofe_highz, chartab_2024_stellar_mzr, stanton_2024_excels_arh, stanton_2024_alpha_enhancement}.
We set the Ne/O abundance to the best-fitting value from our \textsc{pyneb} analysis.
Ultimately, these abundance choices have little impact on the outcome of this analysis, and the same conclusions would hold under a purely solar abundance pattern.
An overview of the main \textsc{cloudy} parameters is given in Table \ref{table:cloudy}.

    \begin{figure*}
        \centering
        \includegraphics[width=0.95\linewidth]{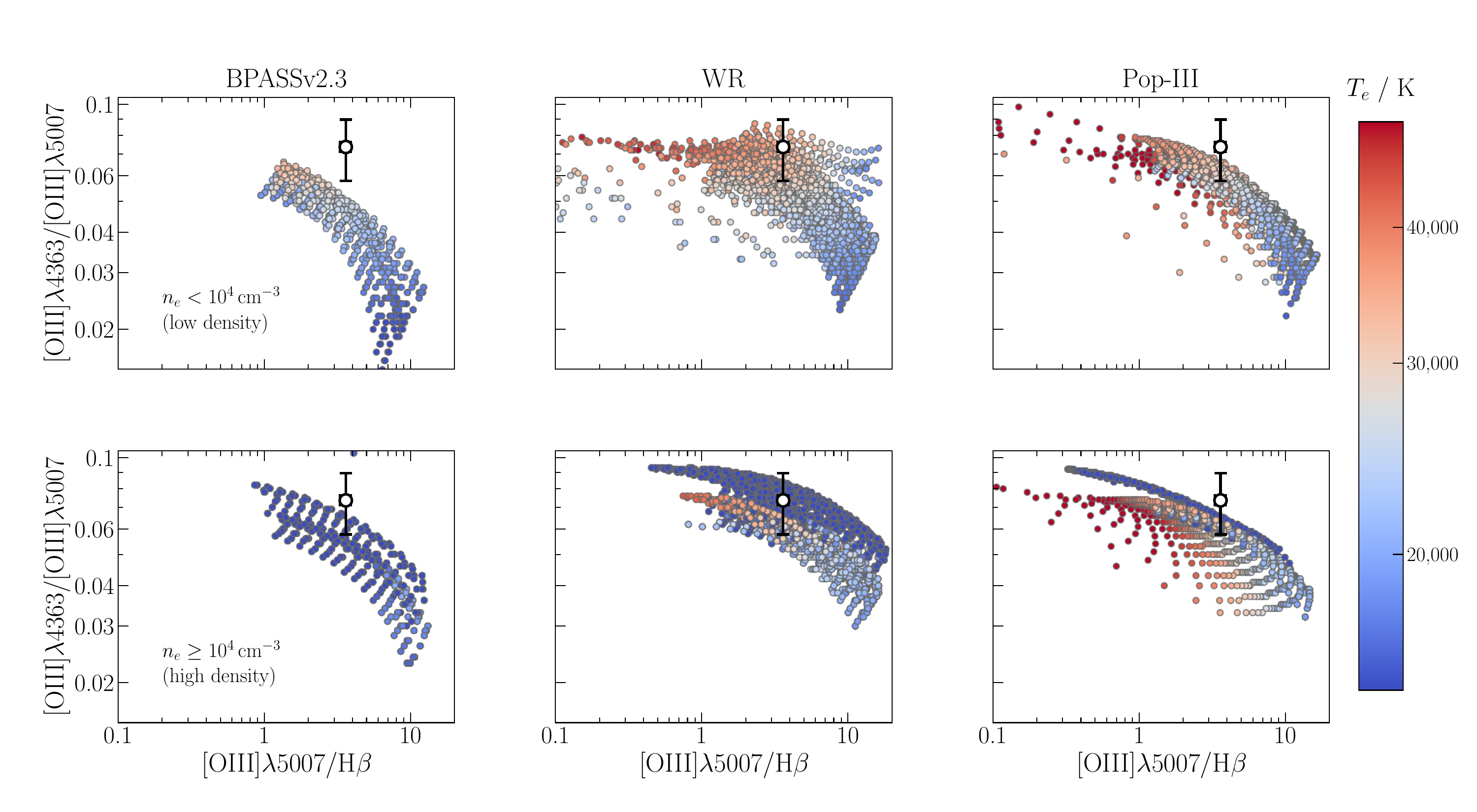}
        \caption{A comparison between the \textsc{cloudy} models and observations in the \oiiia/\hbeta \ versus \oiiiaur/\oiiia \ line ratio diagram. 
        Each column corresponds to the \textsc{cloudy} models associated with one of the three types of stellar SED we consider: BPASSv2.3, Wolf-Rayet (WR) and Pop III (see text for a detailed description of our photoionization modelling framework).
        The top row of panels show the \textsc{cloudy} models in the `low-density' regime which we define to be densities less than $n_e < 10^4 \, \mathrm{cm}^{-3}$.
        The bottom row of panels show the \textsc{cloudy} models in the `high-density' ($n_e \geq 10^4 \, \mathrm{cm}^{-3}$) regime.
        All the \textsc{cloudy} model points are colour-coded by the volume-average ISM electron temperature of the simulated \hii \ region ($T_e$).
        The observed line ratios and associated uncertainties for EXCELS-63107 are shown by the black data point in each panel.
        In the low-density regime, the high \oiiiaur/\oiiia \ we observe require a hot ISM temperature ($\gtrsim 30\,000\, \rm{K}$) that can only be achieved using the WR or Pop III models.
        However, the bottom row of panels demonstrates that the observed line ratios can also be explained by a relatively `cool' ISM temperature at high densities.} 
        \label{fig:cloudy_line_ratios}
    \end{figure*}

\subsubsection{The stellar ionizing continuum spectrum}

The most important factor in determining the ionized gas temperature and resulting nebular spectrum in these models is the choice of the input stellar ionizing continuum. 
The combination of an extremely blue UV continuum slope (Fig.~\ref{fig:beta_muv}), and the potentially high ionized gas temperature (Fig. \ref{fig:dynesty_corner}), suggests that a non-standard stellar population may be present in EXCELS-63107. 
Accordingly, our aim is to explore evidence for an extreme population by adopting the following three different models for the input ionizing spectrum.

\begin{enumerate}

    \item \textbf{BPASSv2.3:} We use the latest version (v2.3) of the Binary Populations and Spectral Synthesis models \citep{eldridge_2017_bpass, stanway_2018_bpass}, specifically adopting the $\alpha$-enhanced variant presented in \citet{byrne_2022_bpass_alpha_enhanced}.
    The BPASS models are `standard' stellar population models that have generally been successful in explaining the ionizing properties of star-forming galaxies at high redshift \citep[e.g.][]{steidel_2016_fuv_optical}.
    We generate these models assuming binary stellar evolution and a \citet{kroupa_imf} IMF with an upper mass cut-off of $300 \, \rm{M}_{\odot}$.
    To ensure the strongest ionizing spectra, we explore a grid of models encompassing the three lowest metallicities, ${Z=[10^{-5}, 10^{-4}, 0.001]}$, and for burst ages of $[1,2,3,4,5]$ Myr.
    Although these models have extremely low metallicities and young ages, if they are capable of reproducing the spectral features in EXCELS-63107 it would suggest that substantial changes are not required in our understanding of the IMF and/or the ionizing spectrum in very low metallicity environments.

    \vspace{0.5mm}
    \item \textbf{Wolf-Rayet (WR):} Our choice of non-standard stellar models is motivated by the recent analysis of a nebular continuum-dominated galaxy at $z=5.9$ \citep{cameron_2024_nebular_cont}. 
    We first consider models from the Potsdam Wolf-Rayet (PoWR) grids presented in \citet{todt_2015_potsdam_wr}.
    Specifically, we use the WNL-H40 grid and adopt the lowest available metallicity (${Z=0.07 \, \rm{Z_{\odot}}}$).
    This metallicity is somewhat higher than our measured value for EXCELS-63107, but it is the closest possible match.
    We select models that cover a range of radii and effective stellar temperatures between $T_{\mathrm{eff}}=63,000\,\rm{K}$ to $T_{\mathrm{eff}}=112,000\,\rm{K}$.
    The minimum initial stellar mass required to become a WR star is $M_{\mathrm{init}} \simeq 25 \, \mathrm{M_{\odot}}$ at solar metallicity (and potentially higher at lower metallicities; \citealp{crowther_wr_stars, shenar_wr_stars}); therefore, if the WR models are successful at reproducing the spectral features of EXCELS-63107, the first order interpretation would be that this is evidence for a top-heavy IMF, with a greater fraction of $M_{\mathrm{init}} > 25 \, \mathrm{M_{\odot}}$ stars forming in low metallicity environments.

    \vspace{0.5mm}
    \item \textbf{Pop III:} Finally, we consider the new evolutionary models of \citet{larkin_2023_pop3_models} for zero-age main-sequence Population III (Pop III) stars.
    We consider a set of models that cover a range of initial masses between $M_{\mathrm{init}}=12\, \mathrm{M_{\odot}}$ and $M_{\mathrm{init}}= 820 \, \mathrm{M_{\odot}}$ and effective temperatures between $T_{\mathrm{eff}}=51,118\,\rm{K}$ and $T_{\mathrm{eff}}=110,629\,\rm{K}$.
    EXCELS-63107 is obviously not a metal-free object, but these models can serve as a general proxy for the harder stellar ionizing spectra which might be required to explain the EXCELS-63107 spectrum.
    It is also worth noting that isolated Pop III formation channels within pre-enriched halos have been identified in some simulations \citep[e.g.][]{correa_magnus_2024_pop3_formation_channel}; it may therefore be possible for Pop III stars to ionize metal-enriched gas in their surroundings, and observing metals in a nebular spectrum may not necessarily rule out Pop III stars as the ionizing source. 
\end{enumerate}

Each stellar model provides the input stellar ionizing continuum spectrum for \textsc{cloudy}. 
We assume that the non-ionizing UV continuum light is dominated by the same population supplying the ionizing photons, and therefore normalise each model to the rest-frame UV photometry.
By doing this, we are inputting the absolute ionizing continuum flux to the \textsc{cloudy} simulations, and the results can be compared against both the line ratios \emph{and} line luminosities.

In total, we consider 15 BPASSv2.3 models, 66 WR models, and 37 Pop III models. 
We run each model across the full range of \hii \ region parameters outlined in Table \ref{table:cloudy}, resulting in a total of 35,400 \textsc{cloudy} simulations.

\subsubsection{The [O\textsc{iii}]$\lambda 4363$/[O\textsc{iii}]$\lambda 5007$ ratio}

We first consider the model predictions for the temperature-sensitive \oiiiaur/\oiiia \ line ratio.
In Fig. \ref{fig:cloudy_line_ratios}, we show a comparison between the models and observations in the \oiiiaur/\oiiia$-$\oiiia/\hbeta \ line ratio diagram, highlighting several important results.
First, from the top row of panels it can be seen that, in the lower density regime ($n_e < 10^4 \, \rm{cm}^{-3}$), it is indeed possible to heat the ionized gas to a volume-averaged temperature of $T_e \gtrsim 35,000 \, \rm{K}$ and match the observed auroral-to-forbidden ratio of \oiiiaur/\oiiia$=0.074$.
Crucially, however, these hot ionized gas temperatures can only be achieved using the WR and Pop III stellar models, which reach the stellar effective temperatures required to provide the necessary heating ($T_{\rm{eff}} \gtrsim 70,000 \, \rm{K}$).
In addition, the gas-phase metallicities of the ISM must be low enough to suppress cooling; we find that high electron temperatures are only a feature of models with low gas-phase metallicities of $Z/\rm{Z}_{\odot} < 0.05$ (which roughly translates to \oiiia/\hbeta \ $\lesssim 4$ for the photoionization models shown in Fig. \ref{fig:cloudy_line_ratios}).

In contrast, the ionizing spectra of the standard BPASSv2.3 models are only capable of heating the gas to ${T_e \simeq 30,000 \, \rm{K}}$.
The resulting \textsc{cloudy} simulations therefore do not provide a good match to the observed \oiiiaur/\oiiia \ ratio (Fig. \ref{fig:cloudy_line_ratios}; see also the comparison to the \citealp{nakajima_2022_photoion_models} models in Fig \ref{fig:nakajima_models}).
Fundamentally, the ionizing continuum spectra of the BPASSv2.3 models are not strong enough to provide the required heating.
This significant difference in the ionizing continuum spectra of the BPASSv2.3, WR and Pop III models is clearly evident in Fig. \ref{fig:spec_and_sed}.
In general, our photoionistaion models suggest that extremely hot gas temperatures of $> 30,000 \rm{K}$ are plausible within $r_e \lesssim 200 \, \rm{pc}$ \hii \ regions, but only by invoking non-standard ionizing sources with effective temperatures of $T_{\rm{eff}} \gtrsim 70,000 \, \rm{K}$.

However, the degeneracy between $T_e$ and $n_e$ discussed above is again demonstrated in the bottom row of panels in Fig. \ref{fig:cloudy_line_ratios}.
In these panels, we only show the models with $n_e \geq 10^4 \, \rm{cm}^{-3}$.
At such high densities, the gas can be efficiently cooled to $T_e < 20,000 \, \rm{K}$ even when heated by an ionizing source with $T_{\rm{eff}} > 70,000 \, \rm{K}$.
At the same time, high \oiiiaur/\oiiia \ ratios are maintained due to the lower critical density of the \oiiia \ transition.
Interestingly, in the high-density scenario, all stellar models perform similarly well, and the preference for the WR or Pop III models is weaker. 
Based solely on this line ratio diagram, there is clearly a degeneracy between the high-temperature and high-density models.
    
    \begin{figure}
        \centering
        \includegraphics[width=\columnwidth]{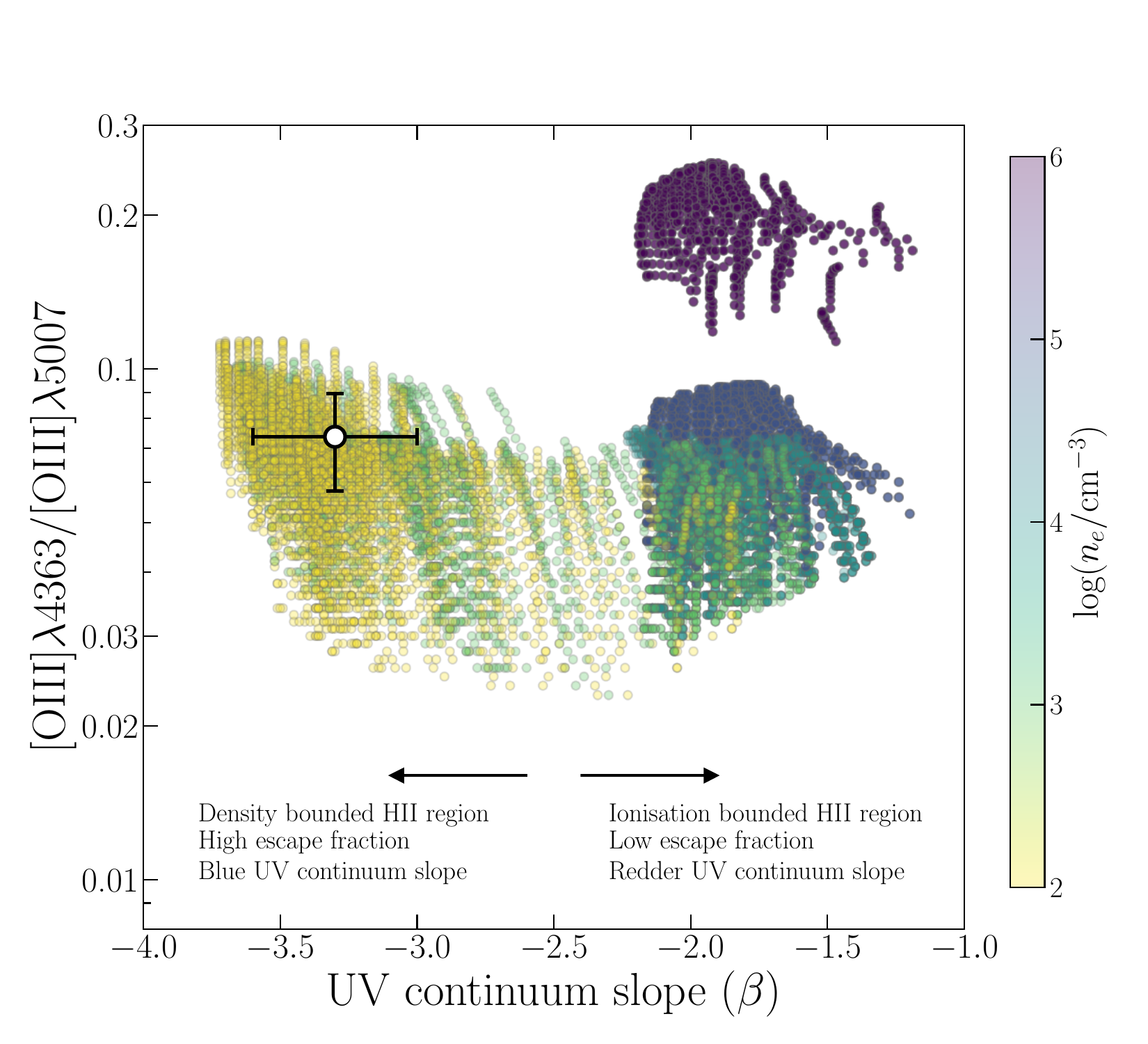}
        \caption{A comparison between the \textsc{cloudy} models and observations in the UV continuum slope ($\beta$) versus \oiiiaur/\oiiia \ diagram.
        The coloured points show the \textsc{cloudy} model outputs for simulation using the WR stellar SEDs; the points are colour-coded by the ISM electron density.
        We only show the outputs for the WR stellar SEDs for clarity but note that the model outputs are similar for the BPASSv2.3 and Pop III SEDs.
        The EXCELS-63107 measurements and associated uncertainties are shown by the black data point.
        A generic feature of our \textsc{cloudy} models is visible.
        At low ISM densities, density-bounded \hii \ regions are more likely to form resulting in a weak nebular continuum and therefore blue UV continuum slope.
        At high densities, compact ionisation-bounded \hii \ regions form and the nebular continuum contribution is substantial, resulting in redder UV slopes (see text for discussion).
        The observations clearly favour a low-density ISM and density-bounded \hii \ region.} 
        \label{fig:cloudy_beta_o3r}
    \end{figure}

    \begin{figure}
        \centering
        \includegraphics[width=\columnwidth]{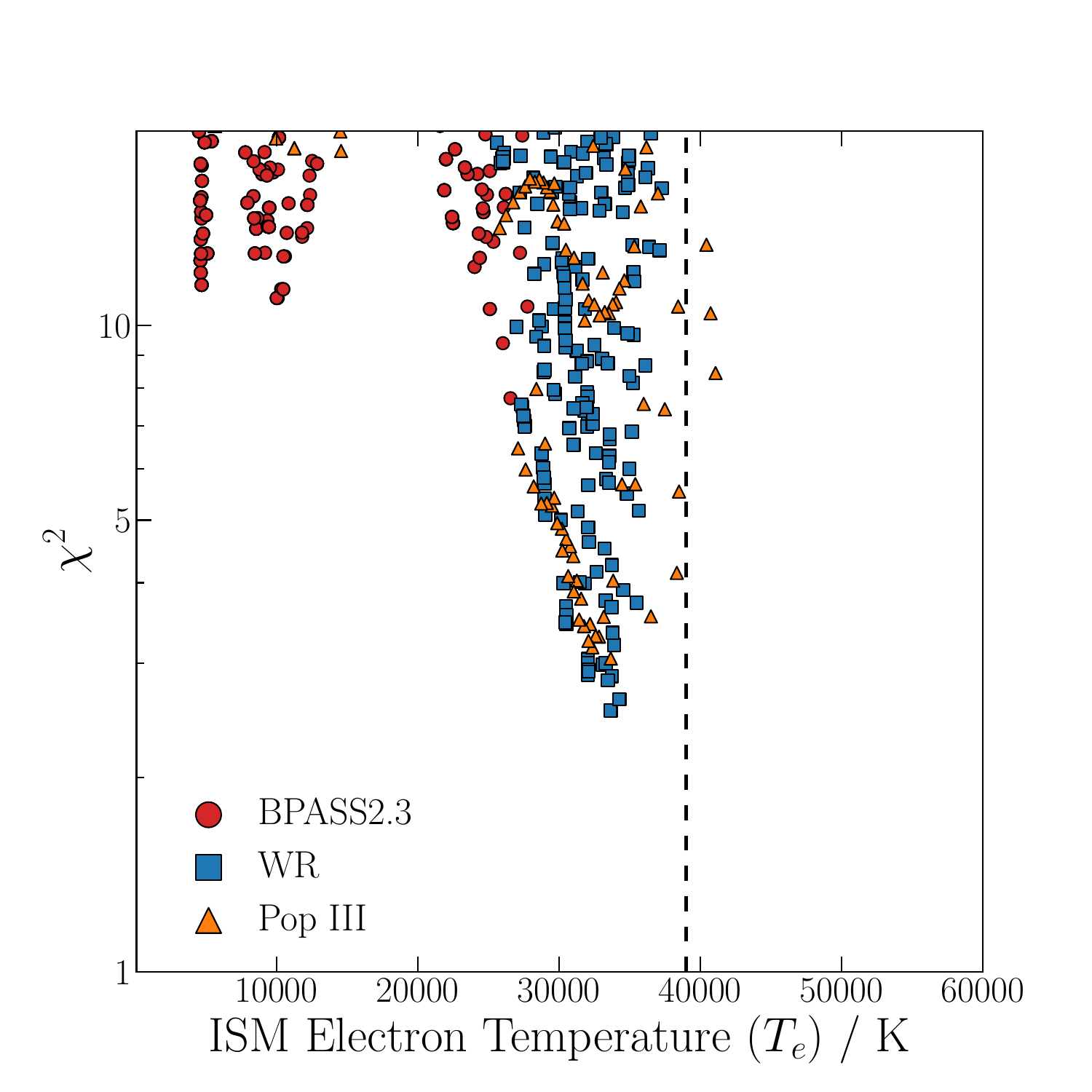}
        \caption{A plot showing the goodness of fit ($\chi^2$) as a function of the volume-averaged ISM electron temperature ($T_e$) for our \textsc{cloudy} models.
        The value of $\chi^2$ is calculated by comparing to a number of observed line ratios, the \oiiia \ line luminosity and the UV continuum slope as described in the text.
        The different colours and symbols show models associated with the three different stellar SEDs as indicated in the legend.
        The vertical dotted lines show the best-fitting value of $T_e$ from the \textsc{pyneb} analysis (see Fig. \ref{fig:dynesty_corner}).
        The best-fitting models are clearly associated with the WR and Pop-III stellar models, and correspond to \hii \ regions with $T_e \simeq 33\,000\,\rm{K}$.
        Full details of the best-fitting models for each of the three stellar SEDs are given in Table \ref{table:best_fit_cloudy_models}.} 
        \label{fig:cloudy_chi2}
    \end{figure}

\subsubsection{Breaking the degeneracy between $T_e$ and $n_e$: the predicted nebular continuum strength}\label{subsec:breaking_te_ne_degeneracy}

Further progress can be made by considering predictions for the \hii \ region geometries (ionization versus density-bounded) as a function of gas density. 
These geometries dictate the strength of the nebular continuum, which in turn influences the predicted UV continuum slope that can be compared with the observed value.

At higher gas densities, the increased recombination rate will lead to the formation of more compact \hii \ regions.
For the five gas densities we consider, ${\mathrm{log}(n_e / \mathrm{cm}^{-3}) = [2, 3, 4, 5, 6]}$, the corresponding average \hii \ region radii are ${\langle r_{\rm{HII}} \rangle = [80, 60, 23, 9, 2] \, \rm{pc}}$.
Therefore, for gas densities $\gtrsim 10^4 \, \mathrm{cm}^{-3}$, we find that compact, ionization bounded, \hii \ regions typically form within the minimum allowed radius of our simulations ($20 \, \rm{pc}$; Table \ref{table:cloudy}).
In this scenario, the escape fraction of ionizing photons is low ($f_{\rm esc} \simeq 0$) and the strength of the nebular continuum will be significant resulting in relatively `red' values of the UV continuum slope, $\beta$ (see recent discussions in \citealp{cullen_2024_uvslope2, katz_2024_balmerjumps}).
In contrast, at lower densities $(n_e \lesssim 10^4 \, \mathrm{cm}^{-3})$ many of the models reach the maximum radius before an ionization edge is formed, resulting in a density-bounded \hii \ region.
In these density-bounded \hii \ regions, the escape fraction can be high, the nebular continuum weak, and the UV slopes blue.

In Fig. \ref{fig:cloudy_beta_o3r} we show our \textsc{cloudy} simulation grid in the $\beta$-\oiiiaur/\oiiia \ plane.
We measure $\beta$ directly from the emergent spectrum predicted by \textsc{cloudy}, adopting the \citet{calzetti_1994_uvwindows} windows.
From Fig. \ref{fig:cloudy_beta_o3r} it can clearly be seen that the extremely blue UV slope of EXCELS-63107 strongly favours the low-density models.
Fundamentally, this is because the UV slope we observe ($\beta = -3.3 \pm 0.3$) is close to the intrinsic slope of our stellar SED models and so, by definition, the nebular continuum contribution must be minimal.
Density-bounded \hii \ regions with $f_{\rm esc} >> 0$ are therefore required to suppress the nebular continuum emission.
Within our \textsc{cloudy} simulation framework, these density-bounded \hii \ regions can only form in a low density ISM ($n_e \lesssim 10^4 \, \mathrm{cm}^{-3}$).
At higher densities ($n_e > 10^4 \, \mathrm{cm}^{-3}$), compact \hii \ regions are formed and the nebular continuum becomes much stronger, resulting in redder values of $\beta$ (we typically find values of $\beta \gtrsim -2.5$; see \citealp{katz_2024_balmerjumps} for an independent analysis that finds similar values of $\beta$ for dense, ionization-bounded, \hii \ regions).

    \begin{table*}
         \centering
         \caption{Details of the best-fitting \textsc{cloudy} photoionization models for each of the three stellar population models.}
         \label{table:best_fit_cloudy_models}
         \renewcommand{\arraystretch}{1.25}
         \begin{tabular}{lrrrr}
             \hline
             \hline
             Parameter & Observed value & \textbf{BPASSv2.3} & \textbf{Wolf-Rayet} & \textbf{Population III} \\
             \hline
             $Z_{\star}/\mathrm{Z}_{\odot}$ & $--$ & $0.001$ & $0.07$ & $\simeq 0$ \\
             $T_e$ & $--$ & $26,568 \, \rm{K}$ & $33,701 \, \rm{K}$ & $33,666 \, \rm{K}$ \\
             $T_{\rm eff}$ & $--$  & $\simeq 50,000 \, \rm{K}$ & $89,125 \, \rm{K}$ & $83,000 \, \rm{K}$ \\
             $r_{\rm{HII}}$ & $--$ & $50 \, \rm{pc}$ & $50 \, \rm{pc}$ & $50 \, \rm{pc}$ \\
             \hii \ region geometry & $--$ & Density-Bounded & Density-Bounded & Density-Bounded \\
             $\mathrm{log}(n_e / \rm{cm}^{-3})$ & $--$ & $3$ & $3$ & $3$ \\
             $f_{\rm esc}^{\rm a}$ & $--$ & $0.02$ & $0.64$ & $0.55$ \\
             $Z/\rm{Z}_{\odot}$ & $--$ & $0.025$ & $0.02$ & $0.02$ \\
             \oiiiaur/\oiiia & $0.076 \pm 0.016$ & $0.05$ & $0.065$ & $0.064$ \\
             \oiiia/\hbeta & $3.61 \pm 0.30$ & $3.64$ & $3.67$ & $3.46$ \\
             \oiiia /\neiiia & $12.4 \pm 2.4$ & $15.2$ & $13.8$ & $12.1$ \\ 
             \oiiia/\oii & $> 15 \, (2\sigma)$ & $\sim 10^4$ &  $\sim 10^5$ & $\sim 10^5$ \\
             $\beta$ & $-3.3 \pm 0.3$ & $-2.7$ &  $-3.1$ & $-3.0$ \\
             $\mathrm{L}_{\rm{[O\textsc{iii}]}}$ & $42.5 \pm 0.1^{\rm b}$ & $42.5$ &  $42.4$ & $42.4$ \\
             $\chi^2$ & $--$ & $7.7$ &  $2.5$ & $3.1$ \\
             $\chi^2_{\nu}$ & $--$ & $1.9$ &  $0.6$ & $0.8$ \\
             \hline
             \multicolumn{5}{l}{$^{\rm a}$ Calculated from the $\tau_{912}$ at the edge of the \hii \ region reported by \textsc{cloudy}.}\\
             \multicolumn{5}{l}{$^{\rm b}$ The uncertainty on $\mathrm{L}_{\rm{[O\textsc{iii}]}}$ has been artificially enlarged (see text)}\\
        \end{tabular}
    \end{table*}

It is important to note that nebular line emission can still be observed in a density-bound scenario in which the nebular continuum is strongly suppressed.
The reason for this is that the nebular lines are always much brighter than the nebular continuum\footnote{The typical equivalent widths of the \oiiia \ line compared to the nebular continuum in the best-fitting \textsc{cloudy} models are $> 1000 \,${\AA}.}.
These lines can remain brighter than the stellar continuum emission even for high escape fractions.
As we shall see in the next section, models that predict a weak nebular continuum and a high escape fraction can still match the observed \oiiia \ line luminosity.

Overall, this comparison with the observed UV continuum slope clearly favours a low-density ISM with $n_e \lesssim 10^4 \, \mathrm{cm}^{-3}$.
In this case, the high-density line ratio predictions shown in the bottom row of Fig. \ref{fig:cloudy_line_ratios} are disfavoured and the large \oiiiaur/\oiiia \ ratio we observe is more likely to be the result of a high ISM temperature powered by an extremely hot ionizing source.

Finally, we note that we do not expect our $\beta$ estimate to be significantly affected by rest-frame UV emission lines contaminating the \emph{JWST}/NIRCam photometry.
Based on the UV line predictions of our best-fitting \textsc{cloudy} models, we find that the UV lines between $1300-3000 \,${\AA} typically bias photometric measurements blue by $\Delta \beta \simeq 0.1$ which would not affect the conclusions drawn from Fig \ref{fig:cloudy_beta_o3r}.
Fundamentally, the broad \emph{JWST}/NIRCam filters we use (with effective widths of $\gtrsim 2000 \,${\AA}) are not strongly sensitive to the rest-frame UV emission lines which, even in the most extreme cases, have equivalent widths of $\lesssim 50 \,${\AA} \citep[e.g.][]{castellano_2024_highew_uv_lines_z12}.
There is also the possibility that the F115W photometry is affected by \lya \ emission, however we find that the derived UV slope is unchanged if we exclude this filter from the fitting.
Nevertheless, future rest-frame UV spectroscopic observations of EXCELS-63107 will be valuable in helping to improve the constraint on the UV continuum slope.
    
\subsubsection{A summary of the photoionization model analysis}

In Fig. \ref{fig:cloudy_chi2}, we show the goodness of fit ($\chi^2$) as a function of the ISM electron temperature ($T_e$) for a selection of the best-fitting individual \textsc{cloudy} models.
In this figure, the $\chi^2$ has been calculated by comparing the following \textsc{cloudy} model outputs to their observed values: \oiiiaur/\oiiia, \oiiia/\hbeta, \neiiia/\oiiia, $\beta$ and $\mathrm{L}_{\rm{[O\textsc{iii}]}}$ (i.e. the \oiiia \ luminosity\footnote{The intrinsic uncertainty on $\mathrm{L}_{\rm{[O\textsc{iii}]}}$ is much lower than the uncertainties on the line ratios and UV slope. 
In the $\chi^2$ analysis we therefore artificially increase the uncertainty on $\mathrm{L}_{\rm{[O\textsc{iii}]}}$ to ensure that the model is not purely fitted to this value (see Table \ref{table:best_fit_cloudy_models}).}).
This figure demonstrates several important points.
First, it can again be seen that the models based on the WR and Pop III SEDs (blue and orange data points) provide the best match to the data, reaching minimum $\chi^2$ values of $\chi^2 = 2.5$ and $3.1$ respectively (Table \ref{table:best_fit_cloudy_models}).
The corresponding reduced $\chi^2$ values are $\chi^2_{\nu} = 0.6$ and $0.8$.
Remarkably, these best-fitting \textsc{cloudy} models essentially directly reproduce the observed line ratios, absolute line luminosities, and the UV continuum slope.
Second, the best-fitting models correspond to ISM electron temperatures of $T_e \simeq 30\,000 - 40\,000 \, \rm{K}$, in good agreement with the best-fitting \textsc{pyneb} value of $T_e \simeq 40\,000 \, \rm{K}$ (we find that the temperature derived from the photoionisation models is, on average, slightly lower).
Finally, the standard \textsc{BPASSv2.3} models clearly provide a poorer match to the data, reaching a minimum $\chi^2$ value of $\chi^2 = 7.7$ (Table \ref{table:best_fit_cloudy_models}).
Fig. \ref{fig:cloudy_chi2} clearly highlights the key conclusions of our analysis: a preference for a hot ISM and a non-standard stellar ionizing source.

In Table \ref{table:best_fit_cloudy_models} we provide specific details on the best-fitting \textsc{cloudy} model for each of the BPASSv2.3, WR and Pop III stellar SEDs. 
We first focus on the overall best-fitting model ($\chi^2=2.5$) which corresponds to a WR stellar ionizing spectrum.
This stellar SED has an effective stellar temperature of ${T_{\rm eff} = 89\,125 \rm{K}}$ and heats a \hii \ region with ${r_{\rm{HII}} \simeq 50 \, \rm{pc}}$ to ${T_e = 33\,701 \rm{K}}$.
The gas-phase metallicity is $Z = 0.02 \, \rm{Z}_{\odot}$ and the \hii \ region is density-bounded (i.e. for this simulation the maximum \hii \ region radius was set to $50 \, \rm{pc}$ and the simulation reached this radius before an ionization edge had formed). 
The resulting nebular continuum spectrum is therefore relatively weak, resulting in a UV slope of $\beta = -3.1$.
The \textsc{cloudy} output shows excellent consistency with the observed line ratios, \oiiia \ luminosity and $\beta$.
The predicted \oiiiaur/\oiiia \ ratio is $0.065$, slightly lower than, but within $1\sigma$ of, the observed value\footnote{We find models that exactly match the observed ratio tend to be very compact density-bounded \hii \ regions which under-predict the line \oiiia \ luminosity.}.

The best-fitting model using the Pop III stellar SEDs is essentially identical across all of the parameters.
For both the WR and Pop III SEDs, the stellar effective temperature is $> 80\,000\,\rm{K}$. 
In general, we find that models that fall within $\Delta \chi^2 = 4$ (i.e. the $2\sigma$ confidence bound) of the best fit have stellar effective temperatures in the range $65\,000\,\mathrm{K} < T_{\mathrm{eff}} < 100\,000\,\mathrm{K}$ with an average value of $T_{\mathrm{eff}} \simeq 80\,000\,\rm{K}$.
The corresponding ISM electron temperatures fall within the range $30\,000\,\mathrm{K} < T_e < 40\,000\,\mathrm{K}$ with an average value of $T_e \simeq 33\,000\,\mathrm{K}$.
In both best-fitting cases, the \hi \ ionization edge forms at a radius $>> 50 \, \rm{pc}$ and therefore these \hii \ regions are density bounded; as a result, a significant fraction of the ionizing photons escape in these models.
The \textsc{cloudy} output reports the optical depth at $912${\AA} at the edge of the \hii \ region, which we convert to escape fractions of $f_{\rm esc}=0.64$ and $f_{\rm esc}=0.55$ for the best-fitting WR and Pop III models, respectively.
Finally, we note that in both of these best-fitting models the predicted strength of the nebular and stellar \heiiopt \ lines is $\lesssim 10$ per cent of \hbeta, and the non-detection of this line in the EXCELS-63017 spectrum is therefore not surprising given the sensitivity of our observations.

The best-fitting \textsc{cloudy} model based on the BPASSv2.3 SEDs has an offset of $\Delta \chi^2 = 5.2$ compared to the overall best fit.
There are a number of reasons why these stellar models are not favoured.
First, the most significant difference between BPASSv2.3 and the WR/Pop III models is a weaker ionizing continuum, which is only capable of heating the ISM to $T_e \simeq 27\,000\,\rm{K}$ and therefore underpredicts the \oiiiaur/\oiiia \ ratio.
Second, the softer BPASSv2.3 ionizing SED also results in fewer $\mathrm{Ne}^{2+}$ ions (ionization potential $41\,\mathrm{eV}$) relative to $\mathrm{O}^{2+}$ ions (ionization potential $35\,\mathrm{eV}$) and as a result, the \oiiia/\neiiia \ ratio is too high.
Finally, because of the weaker ionizing continuum, the \hii \ region for the best-fitting BPASSv2.3 model is `only just' density bounded and the optical depth at the edge of the \hii \ region is large ($f_{\rm esc}=0.02$)\footnote{We actually know, from running a \textsc{cloudy} simulation with the same SED and set of ISM conditions with an upper limit of $150\, \rm{pc}$ on the outer radius, that the \hi \ ionization edge forms at $r = 58 \, \rm{pc}$ for this model.}.
The nebular continuum is therefore strong and predicted UV slope of $\beta=-2.7$ is redder than observed.

It is worth emphasising again here that these BPASSv2.3 models have among the hardest ionizing spectra of currently available `standard' models. 
Moreover, we find that the best-fitting BPASSv2.3 SED has a stellar metallicity of $0.1$ per cent solar and an age of $1\,\rm{Myr}$, which in isolation can also be considered `extreme' compared to the typical stellar populations expected at $z\simeq8$.
Despite this, these ultra low-metallicity, ultra young BPASSv2.3 models are not quite capable of reproducing the observed features of EXCELS-63107.

\begin{figure}
        \centering
        \includegraphics[width=0.85\columnwidth]{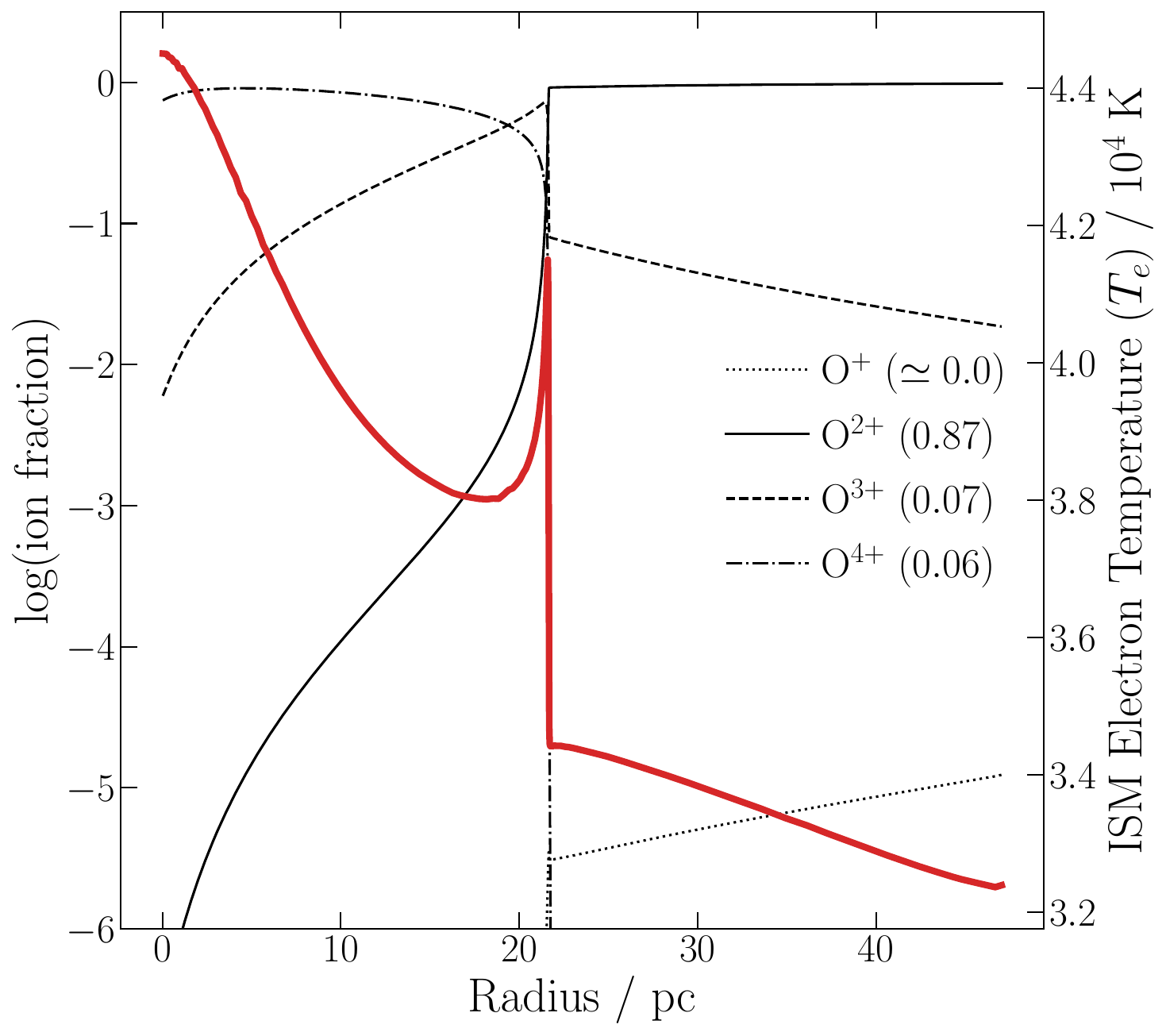}
        \caption{The ionization and temperature structure of the best-fitting \textsc{cloudy} model.
        The \hii \ region shown here is density-bounded with an outer radius of $50 \, \rm{pc}$ and is heated by a WR SED with $T_{\mathrm{eff}} =89\,251\, \rm{K}$ (see Table \ref{table:best_fit_cloudy_models} for full details of the model).
        The black lines show the ionisation fraction of the various ionisation states of O as indicated in the legend; the values are given on the left-hand axis.
        In the legend, the numbers in brackets give the volume-averaged ionization fractions of the various ions.
        For this \hii \ region, $\simeq 87$ per cent of O is in the $\mathrm{O}^{2+}$ state with the remainder in the $\mathrm{O}^{3+}$ and $\mathrm{O}^{4+}$ ionization states.
        The temperature structure is shown by the red line with the temperature values given on the right hand axis.
        The volume-averaged temperature is $T_e = 33\,701\,\rm{K}$.
        The nebular emission spectrum and UV continuum spectrum predicted for this \hii \ region almost exactly matches the observations of EXCELS-63107 (see Table \ref{table:best_fit_cloudy_models}).} 
        \label{fig:cloudy_ionization_structure}
    \end{figure}

It can be seen from Table \ref{table:best_fit_cloudy_models} that the best-fitting models have ISM metallicities of $Z \simeq 0.02 \, \rm{Z}_{\odot}$, which is slightly higher than the metallicities estimated using \textsc{pyneb}, albeit well within the $1\sigma$ uncertainty (see Table \ref{table:physical_params}).
If we consider all models that are within $\Delta \chi^2 = 4$ of the minimum value, the range of metallicities is between $1.6$ per cent and $5$ per cent solar, with an average value of $2.2$ per cent.
One reason why the \textsc{pyneb} analysis might be slightly underestimating the true oxygen abundance is because of our assumption that $\rm{O}^{2+}/H \simeq \rm{O}/H$ (see Section \ref{subsec:pyneb_analysis}).
As it turns out, this assumption is probably not quite correct.
In Fig. \ref{fig:cloudy_ionization_structure} we show the ionisation structure of the best-fitting \textsc{cloudy} model, where it can be seen that a small, but non-zero, fraction of the gas-phase O exists in the ionization states $\rm{O}^{3+}$ and $\rm{O}^{4+}$ .
The volume-averaged fraction of $\rm{O}^{2+}$ in this model is $f(\rm{O}^{2+})=0.87$, therefore, although $\rm{O}^{2+}$ is still the dominant ionisation species, an ionization correction factor of 1.15 would need to be applied in order to convert from the $\rm{O}^{2+}$ abundance to the total O abundance.
Applying this correction factor to the \textsc{pyneb} results increases the metallicity estimate to $Z = 0.016^{+0.011}_{-0.005} \, \rm{Z}_{\odot}$.
Taking the \textsc{pyneb} analysis and photoionization models together, it is reasonable to assume that the true gas-phase metallicity is likely to be somewhere between $1-3$ per cent solar.
Note that the volume ionization fraction of $\mathrm{O}^{+}$ is essentially zero, hence the extremely large predicted \oiiia/\oii \ ratios in Table \ref{table:best_fit_cloudy_models}.

To summarise, the photoionization model analysis agrees with the results of the \textsc{pyneb} forward modelling in that a hot ISM temperature ($T_e > 30\,000\,\rm{K}$) and low metallicity ($Z\simeq0.01-0.03\,\rm{Z}_{\odot}$) is preferred as the most likely explanation for the observed \oiiiaur/\oiiia \ ratio.
Crucially, however, these photoionization models also provide additional physical insights, namely demonstrating that a hot ionizing source with $T_{\mathrm{eff}} \simeq 80\,000\,\rm{K}$ is required to heat the ISM to the required temperature (Figs. \ref{fig:cloudy_beta_o3r} and \ref{fig:cloudy_chi2}), and that the \hii \ region must be density bounded to account for the observed blue UV slope (Fig. \ref{fig:cloudy_beta_o3r}), which implies a high ionizing photon escape fraction ($f_{\rm esc} \simeq 0.5-0.7$).

\section{Discussion}
\label{sec:discussion_of_excels63107}

The observations presented here demonstrate the tremendous progress being made in our understanding of the chemical properties of early galaxies with \emph{JWST}.
It is now clearly possible to measure direct-method metallicities down to at least $Z \simeq 0.01\,\rm{Z}_{\odot}$ in compact star-forming systems at $z>8$.
By reaching these metallicities, we are at a boundary below which no galaxy has been observed at any redshift \citep[e.g.][]{kojima_2020_empgs}, and we are entering a regime in which the physics of star formation and the properties of massive stars become highly uncertain.
Indeed, EXCELS-63107 is the first example of a galaxy with a direct metallicity constraint of $Z \simeq 0.01\,\rm{Z}_{\odot}$ at high redshift, and there are already convincing signs that standard star formation and stellar evolution prescriptions may not apply to this object.
Below, we discuss what our analysis implies about the nature of the massive stellar population in EXCELS-63107, and discuss our metallicity constraint within the context of other $z\simeq8$ galaxies and extremely metal-poor galaxies in the local Universe.

\subsection{The nature of the massive stellar population in EXCELS-63107}
\label{subsec:comparison_to_local}

Our photoionisation model analysis suggests that a standard stellar population combined with a standard IMF is not favoured as an explanation for the observed spectrum and SED of EXCELS-63107.
Instead, the more extreme ionizing continuum spectra supplied by the WR and Pop III models are clearly preferred.
Although these pure WR and pure Pop III models do not directly correspond to a physically realistic scenario, their preference can be interpreted in two ways: as evidence for a top-heavy IMF and/or the presence of an exotic, potentially metal-free stellar population.

\subsubsection{Evidence for a top-heavy IMF?}

The overall best-fitting model is based on the Wolf-Rayet star stellar SEDs, and we first consider what this might imply for the mass distribution of stars, and the possibility of a top-heavy IMF, in EXCELS-63107.
Based on Milky Way observations, the progenitors of Wolf Rayet stars are expected to have a minimum stellar mass of $M = 25 \, \rm{M}_{\odot}$ at solar metallicity \citep{crowther_wr_stars}.
This minimum mass is likely to increase in low-metallicity environments.
For a sample of seven WR stars in the SMC, \citet{hainich_2015_wr_smc} find a mass range of $37 - 75 \, \rm{M}_{\odot}$ with a median of $43 \, \rm{M}_{\odot}$.
Even accounting for the lower mass-loss rates at low metallicities, the progenitor masses of these SMC WR stars are likely to be $\gtrsim 50  \, \rm{M}_{\odot}$.
If we interpret our results as indicating that the stellar population of EXCELS-63107 contains a larger fraction of WR stars compared to typical star-forming galaxies, then the implication is that there must be an excess of $M > 50 \, \rm{M}_{\odot}$ stars compared to a standard IMF.

Interestingly, the low metallicity of EXCELS ($\simeq 0.01 \, \rm {Z}_{\odot}$) places it at a boundary in metallicity where a transition to a top-heavy IMF is expected based on theoretical models.
For example, the high-resolution hydrodynamic simulations of \cite{chon_2021_imf_lowz} predict that the transition between a top-heavy and standard Chabrier-like IMF occurs at a critical metallicity of $Z \simeq 0.01-0.1 \, \rm {Z}_{\odot}$.
In the low-metallicity regime, slower cooling and decreased cloud fragmentation lead to the preferential formation of more massive stars.
This result has also been validated in subsequent works exploring the effects of CMB heating and radiative feedback, both of which further increase the likelihood of massive star formation in low-metallicity environments at high redshifts \citep{chon_2022_cmb_imf, chon2024_feedback_imf}.
The link between the functional form of the IMF and metallicity is physically well motivated \citep[e.g.][]{omukai_2005_lowmetal_imf}, and, as an extremely metal-poor high-redshift galaxy, EXCELS-63107 is exactly the type of object in which a top-heavy IMF is expected.

While our results cannot be directly translated into a constraint on the IMF, we can still make a rough estimate of the potential excess of $\gtrsim 50 \, \rm{M}_{\odot}$ stars.
Following a method similar to one outlined in \citet{cameron_2024_nebular_cont}, we assume that the progenitor WR stars have $M > 50 \, \rm{M}_{\odot}$ such that, for a \citet{kroupa_imf} IMF with an upper mass cutoff of $300 \, \rm{M}_{\odot}$, we would expect $\sim 2 \times 10^4$ WR stars to be present in the stellar population of EXCELS-63107 given our best estimate of the burst mass ($\mathrm{log}(M_{\mathrm{burst}}/\mathrm{M}_{\odot}) \simeq 7.4$; Section \ref{subsec:stellar_mass_estimate})\footnote{The IMF-related numbers in this section are calculated using the python \textsc{imf} library (https://github.com/keflavich/imf/tree/master).}.
However, by scaling the best-fitting WR SED to the EXCELS-63107 photometry (see Fig. \ref{fig:spec_and_sed}) we estimate that $\sim 2 \times 10^5$ WR stars are required to account for the observed UV luminosity (here we are assuming that the WR emission is dominating the non-ionizing UV continuum)\footnote{The luminosity of an individual WR star in the \citet{todt_2015_potsdam_wr} models is $10^{5.3}\mathrm{L}_{\odot}$ but we increase this to $10^{6.2}\mathrm{L}_{\odot}$ which is the typical observed value for low metallicity WR stars \citep[e.g.][]{shenar_wr_stars}.}.
The number of WR stars needed to explain the observations therefore corresponds to a factor $\simeq 10-30 \times$ increase in the number of $> 50 \, \rm{M}_{\odot}$ stars compared to a standard IMF.
Although this is obviously an approximate calculation with significant systematic uncertainties (e.g. the progenitor masses of WR stars, the burst mass of EXCELS-63107, stellar modelling uncertainties, etc.), it nevertheless yields a useful first-order estimate of the massive star excess that might be required to explain the spectrum of EXCELS-63107.
Interestingly, this value is in reasonable agreement with the determination of the high-mass IMF excess required to explain the nebular-dominated galaxy spectrum presented in \citet{cameron_2024_nebular_cont}.

It is worth noting here that, if the IMF is indeed top-heavy, this would affect our derived values of stellar mass and star-formation rate.
Given the already significant uncertainty on the stellar mass based on the unknown star-formation history, we only attempt to derive an approximate correction factor here.
Changing the high-mass power-law slope of the \citet{kroupa_imf} IMF from $-2.3$ to $-1$ yields an approximately $10 \times$ increase in $> 50 \, \mathrm{M}_{\odot}$ stars and results in a $10 \times$ increase in the integrated luminosity of a stellar population at fixed mass \footnote{Based on a simple scaling between the total luminosity of a star its mass \citep{ekstrom_2012_mass_luminosity_relation}.}.
Assuming such a top-heavy IMF could therefore reduce the minimum plausible burst mass by a least an order of magnitude to $\mathrm{log}(M_{\mathrm{burst}}/\mathrm{M}_{\odot}) \lesssim 5.0$.
The star-formation rate would also be reduced by a similar factor.
In principle, the total stellar mass might also reduce significantly, although this assumes that a top-heavy IMF applies to all previous star-formation.
Interestingly, these lower stellar masses would be more compatible with the typical halo masses predicted to host pristine star-formation \citep[e.g.][]{zier_2025_pop3_thesan_zoom}.

    \begin{figure*}
        \centering
        \includegraphics[width=0.7\textwidth]{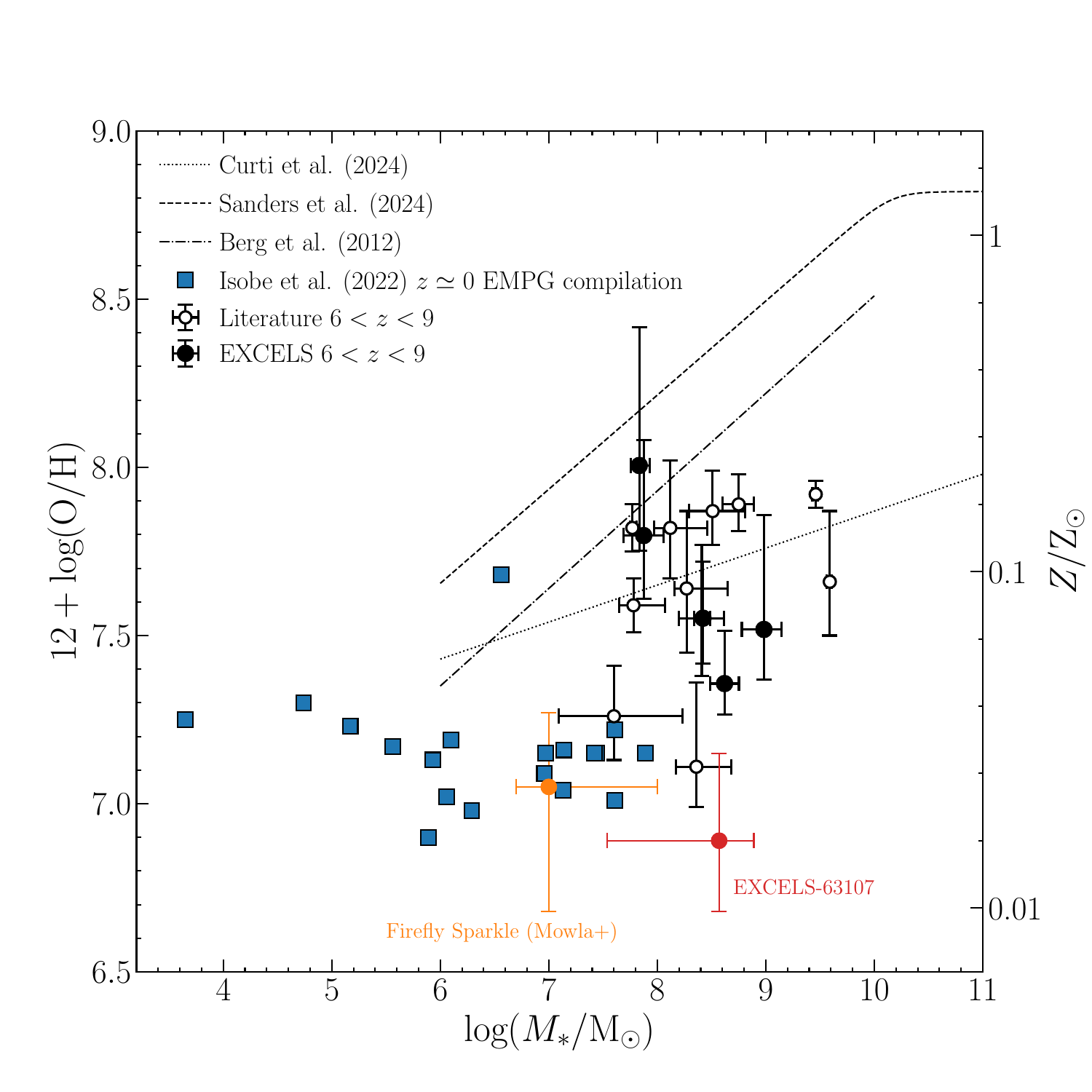}
        \caption{The stellar mass and gas-phase oxygen abundance of EXCELS-63107 compared to measurements from a variety of literature sources.
        All of the sources shown here have been selected to have direct $T_e$-based estimates of the oxygen abundance.
        The red circular data point shows EXCELS-63107 and the orange circular data point shows the `Firefly Sparkle' source from \citet{mowla_2024_firefly}.
        The open black data points show a compilation of literature sources at $6 < z < 9$ and the solid black data points are other galaxies at $6 < z < 9$ from the EXCELS survey that will be presented in a companion paper \citep{scholte_2025_excels}.
        The blue square data points show a compilation of local extremely metal-poor sources from \citet{isobe_2022_local_xmpgs}.
        The dashed and dotted black lines show the mass-metallicity relations at $z=0$ and $z=6-10$ derived from empirical strong-line calibrations from \citet{sanders_2021_mosdef_mzr} and \citet{curti_2024_jades_mzr} respectively.
        The dot-dashed black lines show the mass-metallicity relation for low-luminosity galaxies in the local volume (within $11 \, \rm{Mpc}$ of the Milky Way) with $T_e$-based abundance estimates from \citet{berg_2012_lvl_mzr_z0}.
        At $Z\simeq0.1\,\rm{Z}_{\odot}$, EXCELS-63107 has one of the lowest direct constraints on $\rm{O/H}$ ever observed in a galaxy, and lies on the `metallicity floor' observed for local systems.} 
        \label{fig:mzr}
    \end{figure*}

\subsubsection{Alternative scenarios}

We have also demonstrated that the theoretical Pop III stellar SEDs of \citet{larkin_2023_pop3_models} are compatible with our observations, and one extreme possibility is that a Pop III stellar population has formed within a mildly enriched halo, illuminating the surrounding gas. 
This formation mechanism has recently been proposed as a potential pathway for Pop III star formation, occurring when a sufficiently dense aggregation of pristine gas merges with a pre-enriched halo \citep[e.g.][]{venditti_2023_pop3_pop2_formation, correa_magnus_2024_pop3_formation_channel}. 
While this scenario remains speculative, further investigation into its plausibility at the redshift of EXCELS-63107 ($z = 8.271$) would be valuable.
If this Pop III formation pathway is expected, then searching for the signatures of extremely hot gas in dense, metal-poor star-forming clusters could offer an alternative approach to finding the sites of pristine star formation in the early Universe. 

Very massive stars (VMS) with stellar masses up to $\simeq 1000 \, \rm{M}_{\odot}$ have been proposed to explain extremely broad \heiiuv \ emission in local star clusters \citep[e.g.][]{grafener_2015_vms, wofford_2023_vms} as well as the high $\mathrm{N/O}$ abundances observed in some high-redshift galaxies \citep[e.g.][]{vink_2023_vms}.
Naturally, VMS are expected to produce extremely strong ionizing spectra \citep[e.g.][]{schaeter_2024_vms_imf}.
A VMS scenario would correspond to an extension of the high-mass cutoff of the IMF, which could of course occur simultaneously with a change in the IMF slope \citep{schaeter_2024_vms_imf}.
Unfortunately, to our knowledge, no publicly available VMS models with predictions for the ionizing continuum spectrum are available, and therefore we could not include them in our analysis.

Another plausible candidate may be stripped stars.
Stripped stars with $M \simeq 8-25 \, \rm{M}_{\odot}$ and effective temperatures of ${T_{\rm{eff}} \simeq 50 \, 000 - 100\, 000 \, \rm{K}}$ have been observed in the SMC \citep{gotberg_2023_stripped_stars}.
These stripped stars have luminosities in the range ${\sim 10^4-10^5 \, \rm{L}_{\odot}}$ so that $\sim 10^6-10^7$ such stars would be needed to match the observed UV luminosity of EXCELS-63107.
A standard \citet{kroupa_imf} IMF predicts $\sim 4 \times 10^5$ stripped stars for the stellar mass of EXCELS-63107 and so, in this scenario, the implied excess of massive stars is much greater ($\simeq 10-100 \times$ above a standard IMF).

We have also not considered the possibility of ISM heating from non-thermal sources such as high-mass X-ray binaries (HMXB), which have been explored in other works as an explanation for high \oiiiaur/\oiiia \ ratios \citep[e.g.][]{katz_2023_strongo3aur} and strong \heii \ emission lines in star-forming galaxy spectra \citep[e.g.][]{senchyna_2020_xrb_heii}.
However, as our observations can be reproduced assuming stellar sources, and given the relatively arbitrary choices needed to construct HMXB models (e.g. the black hole stellar mass, accretion disk radius), we do not consider an appeal to non-thermal heating necessary in this instance.

An obvious avenue for making further progress here will be to obtain rest-frame UV spectroscopy for EXCELS-63107 and other galaxies like it.
The very high ionization lines in the rest-frame UV (e.g. \heiiuv) will provide important new constraints on the ionizing spectrum and are crucial for further clarifying the nature of the massive stellar population.
Based on the predictions of our \textsc{cloudy} models, these UV lines should be clearly detected with $\simeq 5-8$ hour exposures with the NIRSpec G235M grating.

\subsubsection{Literature comparison}

Independent evidence for an excess of massive stars in some star-forming galaxies has been found both at high redshifts and in the local Universe.
We have already referred to the recent work of \citet{cameron_2024_nebular_cont} who presented an analysis of a galaxy spectrum dominated by nebular continuum emission at $z=5.943$ (GS-NDG-9422).
The source has an estimated metallicity of $Z \simeq 0.08 \, \rm{Z}_{\odot}$ and \citet{cameron_2024_nebular_cont} argue that the Balmer jump and prominent UV continuum turnover observed in the NIRSpec/PRISM spectrum can be explained by extremely hot stars with $T_{\rm eff} \sim 10^5 \, \rm{K}$ that provide sufficient ionizing photons to produce a visible two-photon continuum.
As discussed above, \citet{cameron_2024_nebular_cont} also estimate that compared to a standard IMF, a factor $\simeq 10 \times$ increase in the number of $> 50 \, \rm{M}_{\odot}$ stars is needed to reproduce the GS-NDG-9422 spectrum.
Although the specific physical scenarios are different, the line of reasoning is therefore very similar.
As an aside, it is interesting to note that our models predict that EXCELS-63107 would be a strongly nebular-dominated galaxy similar to GS-NDG-9422 if the \hii \ region was ionization-bounded and the escape fraction was $\simeq 0$.

Following on from the \citet{cameron_2024_nebular_cont} result, \citet{katz_2024_balmerjumps} have found five additional nebular-dominated galaxy candidates at $4 < z < 6$ in public NIRSpec/PRISM data, and also favour a top-heavy IMF as the most likely physical interpretation of their spectra (see also \citealp{saxena_2024_uvslopes}).
However, it is important to note that other explanations for these UV turnovers have been proposed, including strong damped \lya \ systems \citep[DLA; e.g.][]{heintz_2024_dla_nebcont_debate, terp_2024_dla_nebcont_debate} or offset AGN emission combined with a DLA \citep[e.g.][]{tacchella_2024_agn_nebcont_debate, li_2024_agn_nebcont_debate}.
Nevertheless, it is clear that a top-heavy IMF is one plausible explanation for these objects.

Observations of the gravitatioanlly-lensed Sunburst Arc at $z=2.37$ with \emph{JWST}/NIRSpec Integral Field Unit (IFU) have revealed broad stellar \heiiopt \ emission in one compact LyC emitting cluster combined with enhanced nitrogen enrichment (\citealp{rivera-thorsen-sunburst-wr-features}; see also \citealp{pascale_2023_sunburst_arc_no} and \citealp{welch_2025_sunburst_arc_abundances}).
These observations are direct evidence for a significant WR population, and, as discussed by these authors, potentially hint at a top-heavy IMF in this cluster.

A more directly comparable result to our own comes from the NIRSpec/PRISM observations of a lensed galaxy at $z=8.3$ (Firefly Sparkle) presented in \citet{mowla_2024_firefly}.
The Firefly Sparkle is lensed by a factor $\simeq 20$, and the reconstructed source plane image reveals that it has a half-light size of $r_e \simeq 300 \, \rm{pc}$ and is composed of ten distinct star-forming clumps.
The NIRSpec/PRISM spectrum (covering a number of the individual clumps) shows a strong blended \hgamma \ plus \oiiiaur \ emission feature and a Balmer jump that the authors attribute to a hot electron temperature due to top-heavy IMF (i.e. a similar argument to the one we are making).
The \citet{mowla_2024_firefly} object is also estimated to be very metal-poor ($Z \simeq 0.02 \, \rm{Z}_{\odot}$) although it differs from EXCELS-63107 in that it has a relatively red UV continuum slope due to strong nebular continuum emission.
Overall, given the heavily-blended \oiiiaur \ line in their spectrum, the resulting \oiiiaur/\oiiia \ ratio, and hence the determination of $T_e$, is less robust than for EXCELS-63107.
However, the existence of EXCELS-63107 supports their interpretation.
Likewise, the source-plane image reconstruction presented in \citet{mowla_2024_firefly} also provides useful information about the likely physical structure of EXCELS-63107.

Our analysis and the examples described above fall into a distinct category in which a top-heavy IMF is required to explain specific spectral features (e.g. a large \oiiiaur/\oiiia \ ratio or a nebular-dominated spectrum with prominent UV downturn).
Other studies have invoked a top-heavy IMF to explain chemical abundance ratios, specifically of the CNO elements.
These include the $^{13}\rm{C}/^{18}\rm{O}$ isotope ratio in starburst galaxies at $z=2-3$ \citep{zhang_2018_th_imf} and the large $\rm{N}/\rm{O}$ abundance ratios observed in a number of metal-poor high redshift objects \citep{bekki_2023_th_imf_no_ratios, curti_2024_z9_cno}.
However, in a companion EXCELS survey paper we find that an excess of massive stars is not necessarily required to explain the CNO abundances of two star-forming galaxies at $z\simeq5$ \citep{arellano_cordova_2024_excels_cno}. 
Crucially, these two sources are relatively metal-rich ($Z \simeq 0.2-0.3 \, \rm{Z}_{\odot}$) and outside of the metallicity regime where top-heavy IMFs are expected theoretically.
Several studies have also invoked a top-heavy IMF to explain the evolution of the UV luminosity function at high redshifts \citep[e.g.][]{cueto_2024_th_imf_uvlf, hutter_2024_th_imf_uvlf, lu_2025_galform_th_imf}.
In the local Universe, more direct evidence for a top-heavy IMF has been presented by \citet{schneider_2018_30dor_th_imf} and \citet{kalari_2018_ngc796_th_imf} in the 30 Doradus and NGC 796 young star-forming clusters, respectively, although the inferred massive star excess in these cases is much less dramatic (for example \citealp{schneider_2018_30dor_th_imf} infer a factor $1.3 \times$ increase in the number of stars with $> 30 \, \rm{M}_{\odot}$ compared to a \citealp{salpeter_imf} IMF).

Our observations and those of \citet{mowla_2024_firefly} demonstrate a new way in which variations in the IMF can be explored at high redshift using a direct determination of the ISM electron temperature in extremely metal-poor systems. 
A systematic analysis with a much larger sample size should be possible with dedicated \emph{JWST} observations.

\subsection{The typical metallicities of galaxies at $\mathbf{z\simeq8}$ and a comparison to local metal-poor systems}
\label{sec:mzr}

In Fig. \ref{fig:mzr}  we compare the stellar mass and metallicity of EXCELS-63107 with other galaxies at $6 < z < 9$ that have temperature-based oxygen abundance estimates (taken from \citealp{nakajima_2023_z410_mzr}, \citealp{laseter_2024_direct_oh_jades}, \citealp{morishita_2024_direct_oh_highz}, \citealp{mowla_2024_firefly} and \citealp{sanders_2024_jwst_highz_te}).
We also include five more galaxies at $6 < z < 8$ from the EXCELS survey that will be presented in a companion paper \citep{scholte_2025_excels}.
In addition, we show a selection of the most metal-poor systems uncovered in the local Universe from \citet{isobe_2022_local_xmpgs}.
The \citet{isobe_2022_local_xmpgs} compilation was obtained from a variety of different sources \citep{izotov_1998_local_xmpg, izotov_2009_local_xmpg, izotov_2018_local_xmpg, izotov_2019_local_xmpg, izotov_2021_local_xmpg, skillman_2013_local_xmpg, hirschauer_2016_local_xmpg, saachi_2016_local_xmpg, hsyu_2017_local_xmpg,  senchyna_2019_local_xmpg, kojima_2020_empgs}. 

The first point to note is that EXCELS-63107 has one of the lowest direct-method oxygen abundances measured in any galaxy to date.
When applying the ionization correction factor suggested by our photoionisation models, we find a best-fitting $T_e$-based value of ${\mathrm{12+log(O/H)}=6.89^{+\,0.26}_{-0.21}}$ which compares to ${\mathrm{12+log(O/H)}=6.89\pm0.03}$ for the lowest value reported in the local Universe from \citet{kojima_2020_empgs}.
Therefore, although we are at the boundary of any directly-inferred oxygen abundance ever observed, it is important to note that metallicities of the order $\simeq 1$ per cent solar are not unprecedented in the local Universe.

However, as can be seen from Fig. \ref{fig:mzr}, almost all of the local sources have a lower stellar masses.
For example, the \citet{kojima_2020_empgs} object has $\mathrm{log}(M_{\star}/\mathrm{M}_{\odot})=5.89^{+\,0.10}_{-0.09}$.
Interestingly, none of these local objects display \oiiiaur/\oiiia \ ratios as large as EXCELS-63107, and typically do not show signs of extreme ionisation conditions (i.e. the typical \oiiia/\oii \ ratios are not particularly high).
One exception to this is the the dwarf galaxy J1046+4047 presented by \citet{izotov_2024_higho32}, which has \oiiia/\oii$=57$ and $\mathrm{12+log(O/H)}=7.08$.
The electron temperature of this object is high ($T_e \simeq 25 \, 000 \, \rm{K}$) but not as extreme as EXCELS-63017; however, this object is probably the closet local analogue, albeit with a much lower stellar mass ($M_{\star} = 1.8 \times 10^6 \, \mathrm{M}_{\odot}$).
Overall, despite having similar metallicities, the stellar populations in the local and high-redshift metal-poor galaxies appear to be, on average, quite different.
Assuming the EXCELS-63107 spectrum is a result of a top-heavy IMF, this might imply that low metallicity is a necessary but not sufficient condition for the preferential formation of massive stars.
An in-depth comparison with low-metallicity local sources (including others not discussed in detail here; e.g., \citealp{berg_2012_lvl_mzr_z0}; \citealp{mcquinn_2020_local_xmpg}) is beyond the scope of this work, but such a comparison is clearly motivated, especially as we increase the sample size of $Z < 0.1 \, \rm{Z}_{\odot}$ galaxies at high redshifts.

The comparison with other $6 < z < 9$ galaxies in Fig. \ref{fig:mzr} illustrates the fact that EXCELS-63107 is metal poor relative to typical star-forming galaxies at similar redshifts and stellar mass, which tend to have $Z \simeq 0.1 \, \rm{Z}_{\odot}$ (Fig. \ref{fig:mzr}).
Despite the fact that the current sample sizes are relatively small, it is reasonable to assume that EXCELS-63107 is probably $\simeq 10 \, \times$ more metal poor than is typical for star-forming galaxies at $z\simeq8$.
Fig. \ref{fig:mzr} also suggests that the range of metallicities of galaxies in the reionisation epoch might be quite substantial, and that the frequency of $Z \simeq 0.01 \, \rm{Z}_{\odot}$ systems could be relatively large ($\simeq 10$ per cent).
Indeed, this would be consistent with the predictions of some theoretical models \citep[e.g.][]{ucci_2023_astreaus_mzr}.

Finally, it is interesting to note that, based on our estimated stellar mass and star-formation rate for EXCELS-63107, the locally-derived fundamental metallicity relations (FMR) of \citet{andrews_and_martini_z0mzr} and \citet{curti_2020_te_fmr} predict oxygen abundances of $\mathrm{12+log(O/H)}=7.96$ and $8.11$ respectively.
While it is again important to acknowledge the significant systematic uncertainties associated with the estimated stellar mass and star-formation rate, at face value this implies that EXCELS-63017 is offset by $\simeq 1$ dex from the local FMR.
This offset is consistent with general trends seen in large samples at similar redshifts \cite[e.g.][]{curti_2024_jades_mzr, scholte_2025_excels}, and suggests significant differences in the properties of gas inflows and outflows at early epochs.
However, larger samples of $T_e$-based abundance determinations are still needed to infer the true form and scatter of the mass-metallicity and fundamental metallicity relations at $z\simeq8$.
 
In the future, dedicated spectroscopy of objects similar to EXCELS-63107 (i.e. compact with an ultra-blue UV continuum) offers a promising path toward characterising the most metal-poor systems in the early Universe and determining whether objects below the local metallicity floor exist ($Z \simeq 0.01 \, \rm{Z}_{\odot}$; \citealp{mcquinn_2020_local_xmpg}).
It is interesting that the two lowest directly determined metallicities so far fall precisely at this floor, which may be telling us something about enrichment timescales and yields in the early stages of star formation.
On the other hand, \citet{vanzella_pop3_like_cluster} present convincing evidence for gas with $Z < 0.01 \, \rm{Z}_{\odot}$ in a highly-lensed $M \lesssim 10^4 \, \rm{M}_{\odot}$ star cluster at $z=6.639$ based on an extremely low \oiiia/\hbeta \ ratio; whether more massive systems with $Z < 0.01 \, \rm{Z}_{\odot}$ exist remains unclear, but if they do, the example of EXCELS-63107 suggests that they can in principle be detected and their oxygen abundance can be directly inferred with \emph{JWST}/NIRSpec.

\section{Summary and Conclusions}
\label{sec:summary_and_conclusions}

We have presented a detailed analysis of the spectrum of EXCELS-63107, a star-forming galaxy at $z=8.271$ observed as part of the EXCELS survey \citep{carnall_2024_excels}.
The galaxy is notable for its compact morphology, extremely steep UV continuum slope, and strong \oiiiaur \ line emission in the NIRSpec/G395M spectrum (Fig. \ref{fig:spec_and_sed}).
Taken together, our analysis reveals that EXCELS-63107 is one of the most metal-poor star-forming systems ever observed and may host a population of unusual massive stars.
Our main results can be summarized as follows:
\begin{enumerate}

\item From \emph{JWST} PRIMER NIRCam imaging data we infer an extremely compact morphology and blue UV SED. 
We find that EXCELS-63107 is consistent with being unresolved in all \emph{JWST}/NIRCam imaging filters and place a conservative upper limit on the half-light radius of $r_e < 200 \, \rm{pc}$ (Fig. \ref{fig:galfit}).
Given the compact morphology, we hypothesise that EXCELS-63107 is a single star-cluster complex (or giant \hii \ region) similar in scale to 30 Doradus and IIZw40. 

\item We measure a UV continuum slope of $\beta = -3.3 \pm 0.3$ which is much bluer than is typical for galaxies with the same UV magnitude (Fig. \ref{fig:beta_muv}), and is consistent with a dust-free stellar continuum in the absence of a strong nebular continuum contribution \citep[e.g.][]{cullen_2024_uvslope2}.
The $\hgamma/\hbeta$ and $\hdelta/\hbeta$ Balmer line ratios measured from the NIRSpec/G395M spectrum are also consistent with their theoretical, dust-free values.
We conclude that both the stellar and nebular emission are unattenuated by dust.

\item The NIRSpec/G395M spectrum (Fig. \ref{fig:spec_and_sed}) is characterized by three main features: a large \oiiia/\oii \ ratio suggestive of a density bounded \hii \ region; a large \oiiiaur/\oiiia \ ratio consistent with either a very hot or dense ISM, and a relatively small \oiiia/\hbeta \ ratio indicating a low gas-phase oxygen abundance (see Table \ref{table:line_fluxes}).
Via a forward modelling analysis using \textsc{pyneb} - and accounting for unseen ionisation states - we infer a direct-method oxygen abundance of $12+\mathrm{log(O/H)}$ = $6.89^{+0.26}_{-0.21}$ (i.e. $\simeq 1.6$ per cent of the solar value) making EXCELS-63107 one of the lowest metallicity galaxies observed to date.

\item More remarkably, the \textsc{pyneb} analysis, fundamentally driven by the large \oiiiaur/\oiiia \ ratio, yields a best-fitting ISM electron temperature of $T_e \simeq 40 \, 000 \, \rm{K}$, significantly hotter than anything previously observed.
The high electron temperature also naturally implies an extremely hot ionizing source.
However, we find that the line data alone cannot strongly rule out the alternative scenario in which the ISM is extremely dense ($n_e > 10^4 \, \mathrm{cm}^{-3}$) and the electron temperature is within the normal range (Fig. \ref{fig:dynesty_corner}).

\item To distinguish between these two alternatives, we conduct a detailed photoionization modelling analysis using \textsc{cloudy}, testing a variety of stellar ionizing spectra and \hii \ region geometries (see Section \ref{subsec:cloudy_analysis} for full details).
We find that high-density models are not favoured as they produce extremely compact ($< 50 \, \rm{pc}$) ionization-bounded \hii \ regions that result in strong nebular continuum emission which is inconsistent with the observed UV continuum slope (Fig. \ref{fig:cloudy_beta_o3r}).
Conversely, models with hot ISM temperatures generally result in density-bounded \hii \ regions and are consistent with the UV slope and the large \oiiiaur/\oiiia \ ratio (Figs. \ref{fig:cloudy_line_ratios} and \ref{fig:cloudy_beta_o3r}).

\item The photoionization model which provides the best match to the observed properties of EXCELS-63107 corresponds to a pure Wolf-Rayet star SED with ${T_{\rm eff} = 89\,125 \rm{K}}$.
The resulting \hii \ region has a radius of ${r_{\rm{HII}} \simeq 50 \, \rm{pc}}$ and is heated to a volume-averaged temperature of ${T_e = 33\,701 \rm{K}}$.
The gas-phase metallicity of the best-fitting model is $Z = 0.02 \, \rm{Z}_{\odot}$ (consistent with our \textsc{pyneb} inference) and the \hii \ region is density-bounded with a high implied ionizing photon escape fraction of $f_{\rm esc} = 0.64$.
This model provides an excellent match to the observed line ratios, UV continuum slope and \oiiia \ luminosity ($\chi^2=1.2$; see Table \ref{table:best_fit_cloudy_models}).
The temperature and ionization structure of this \hii \ region is shown in Fig. \ref{fig:cloudy_ionization_structure}.

\item Crucially, we find that the standard BPASSv2.3 stellar population models cannot successfully reproduce the observed spectrum.
Although we have chosen models with an upper mass IMF cutoff of $300 \, \rm{M}_{\odot}$, binary stellar evolution, ages $<5$ Myr and stellar metallicities as low at $\simeq 0.1$ per cent solar, we find that the ionizing SEDs are not capable of heating the ISM to the temperature required to yield the observed \oiiiaur/\oiiia \ ratio (Fig. \ref{fig:cloudy_line_ratios}).
The best-fitting photoionization model based on the BPASSv2.3 SEDs has $\chi^2=2.5$ (Table \ref{table:best_fit_cloudy_models}).

\item Overall, we find that the best-fitting photoionization models correspond to \hii \ regions with ${T_e \simeq 34 \, 000 \rm{K}}$ which requires a non-standard stellar ionizing source with ${T_{\rm eff} > 80\,000 \rm{K}}$ (Fig. \ref{fig:cloudy_chi2}).
In Section \ref{sec:discussion_of_excels63107}, we speculate that this could be explained by a top-heavy IMF resulting in a larger fraction of Wolf-Rayet stars in the stellar population.
A transition to a top-heavy IMF is expected at approximately the metallicity of EXCELS-63107 (i.e. $Z \simeq 0.01 \, \rm{Z}_{\odot}$; \citealp{chon_2021_imf_lowz}).
A rough calculation suggests that the required ionizing continuum could be achieved with a factor $\simeq 10-30$ increase in the number $M > 50 \, \rm{M}_{\odot}$ stars which is consistent with the massive star excess inferred for nebular-continuum dominated spectra at high redshift \citep{cameron_2024_nebular_cont}. 
However, more exotic scenarios, such as pristine star-formation within a mildly enriched halo, are another potential (albeit more speculative) explanation.

\end{enumerate}

Although further observations are ultimately needed to definitively confirm the nature of the stellar population in EXCELS-63107, our current analysis clearly favours an ionizing source that is hotter than those incorporated in commonly-used stellar models combined with a standard IMF.
This is perhaps unsurprising given the fact that EXCELS-63107 is one of bluest, most metal-poor galaxies yet discovered.
Fortunately, if more objects similar to EXCELS-63107 can be uncovered, the prospects for progress are good; our exposure time is relatively modest ($\simeq 2.7$ hours), and additional crucial information can be obtained by observing the rest-frame UV using the shorter-wavelength NIRSpec gratings.
The observations presented here clearly demonstrate that deep medium-resolution spectroscopy of ultra-blue sources is a promising avenue for uncovering extreme stellar populations in the most metal-poor galaxies at cosmic dawn. 

\section*{Acknowledgements}
F. Cullen, K. Z. Arellano-C\'ordova, T. M. Stanton and D. Scholte acknowledge support from a UKRI Frontier Research Guarantee Grant (PI Cullen; grant reference EP/X021025/1). A. C. Carnall, H.-H. Leung and S. Stevenson acknowledge support from a UKRI Frontier Research Guarantee Grant (PI Carnall; grant reference EP/Y037065/1).
J. S. Dunlop thanks the Royal Society for support via a Research Professorship. R. Begley acknowledges the support of the Science and Technology Facilities Council.
This work is based on observations made with the NASA/ESA/CSA James Webb Space Telescope. The data were obtained from the Mikulski Archive for Space Telescopes at the Space Telescope Science Institute, which is operated by the Association of Universities for Research in Astronomy, Inc., under NASA contract NAS 5-03127 for JWST. These observations are associated with program 3543.
For the purpose of open access, the author has applied a Creative Commons Attribution (CC BY) licence to any Author Accepted Manuscript version arising from this submission.

\section*{Data Availability}

All \emph{JWST} and \emph{HST} data products are available via the Mikulski Archive for Space Telescopes (\url{https://mast.stsci.edu}). 
Additional data products are available from the authors upon reasonable request.



\bibliographystyle{mnras}
\bibliography{excels_xmp} 



\appendix


\bsp	
\label{lastpage}
\end{document}